\documentclass[a4paper,12pt]{article}
\usepackage{amsmath}
\usepackage{amssymb}
\usepackage{fullpage}
\usepackage{verbatim}

\usepackage[T1]{fontenc}

\usepackage{graphicx}

\allowdisplaybreaks[1]

\usepackage[hang]{footmisc}
\usepackage[compress,square,numbers]{natbib}
\usepackage[colorlinks,linktocpage,linkcolor=blue,citecolor=blue,urlcolor=blue]{hyperref}

\renewcommand{\d}{\mathrm{d}}
\newcommand{\e}{\mathrm{e}}
\newcommand{\w}{\wedge}

\newcommand{\nl}{\notag \\ &\quad\,}
\newcommand{\nll}{\notag \\ &}

\usepackage{color}

\begin{document}

\numberwithin{equation}{section}

\thispagestyle{empty}

\begin{flushright}
\small MAD-TH-13-07\\
\normalsize
\end{flushright}
\vspace{1cm}

\begin{center}

{\LARGE \bf Brane Curvature Corrections to the}

\vspace{0.5cm}

{\LARGE \bf $\mathcal{N}=1$ Type II/F-theory Effective Action}

\vspace{1cm}
{\large Daniel Junghans$^1$ and Gary Shiu$^{1,2}$}\\

\vspace{1cm}
$^1$ Center for Fundamental Physics \& Institute for Advanced Study,\\
The Hong Kong University of Science and Technology, Hong Kong 

\vspace{0.5cm}

$^2$ Department of Physics, University of Wisconsin,\\
Madison, WI 53706, USA

\vspace{1cm}

{\upshape\ttfamily daniel@ust.hk, shiu@physics.wisc.edu}\\

\vspace{1cm}
\begin{abstract}
We initiate a study of corrections to the K{\"{a}}hler potential of $\mathcal{N}=1$ type II/F-theory compactifications that arise from curvature terms in the action of D-branes and orientifold planes.
We first show that a recently proposed correction to the K{\"{a}}hler coordinates, which was argued to appear at order $\alpha^{\prime 2}g_s$ and be proportional to the intersection volume of D$7$-branes and O$7$-planes, is an artifact of an inconvenient field basis in the dual M-theory frame and can be removed by a field redefinition of the 11D metric. We then analyze to what extent curvature terms in the DBI and WZ action may still lead to corrections of a similar kind and identify two general mechanisms that can potentially modify the volume dependence of the K{\"{a}}hler potential in the presence of D-branes and O-planes. The first mechanism is related to an induced Einstein-Hilbert term on warped brane worldvolumes, which leads to a shift in the classical volume of the compactification manifold. The resulting corrections are generic and can appear at one-loop order on branes and O-planes of various dimensions and for configurations with or without intersections. We discuss in detail the example of intersecting D$7$-branes/O$7$-planes, where a correction can appear already at order $\alpha^{\prime 2}g_s^2$ in the K{\"{a}}hler potential. Due to an extended no-scale structure, however, it is then still subleading in the scalar potential. We also discuss a second mechanism, which is due to an induced D$3$-brane charge in the WZ action of D$7$-branes. Contrary to the first type of corrections, it appears at open string tree-level and shifts the definition of the K{\"{a}}hler coordinates in terms of the classical volume but leaves the volume itself uncorrected. Our work has implications for moduli stabilization and model building and suggests interesting generalizations to F-theory.
\end{abstract}

\end{center}

\newpage
\tableofcontents
\vspace{0.5cm}

\section{Introduction}

The extended nature of strings is what distinguishes string theory from other ultraviolet completions of general relativity. Yet, except for some simple cases, our present knowledge of string theory is often limited
to the small corner of perturbative string theory at large volume and small string coupling, where the theory is well-approximated by classical two-derivative supergravity supplemented by its leading corrections (organized as a double expansion in higher derivatives $\alpha'$ and string loops $g_s$). The existence of a limit where these perturbative corrections are under parametric control is essential for calculation reliability and is assumed in practically all known mechanisms to obtain string compactifications with stabilized moduli. These include mechanisms that employ both perturbative and non-perturbative effects such as 
type IIB/F-theory moduli stabilization along the lines of the KKLT scenario \cite{Kachru:2003aw}, the large volume scenario \cite{Balasubramanian:2005zx, Conlon:2005ki} or the K{\"{a}}hler uplifting scenario \cite{Balasubramanian:2004uy, Westphal:2006tn, Rummel:2011cd, Louis:2012nb}. The possibility of stabilized vacua using only perturbative corrections was investigated in \cite{Berg:2005yu, Green:2011cn, Gautason:2012tb, Dasgupta:2014pma}. 
Furthermore, there have been attempts to construct stabilized compactifications within classical supergravity (see, e.g., \cite{Derendinger:2004jn, Villadoro:2005cu, Camara:2005dc, DeWolfe:2005uu, Silverstein:2007ac, Caviezel:2008ik, Caviezel:2008tf, Flauger:2008ad, Haque:2008jz, Danielsson:2009ff, Danielsson:2011au}).

Although some $\alpha^\prime$ and $g_s$ corrections to the 4D effective action of $\mathcal{N}=1$ type IIB compactifications have been derived in recent years \cite{Becker:2002nn, Berg:2004ek, Berg:2005ja, Berg:2007wt, Berg:2011ij, Cicoli:2007xp, Anguelova:2010ed}, a complete understanding even of the leading corrections is still lacking. The same is true for $\mathcal{N}=1$ F-theory compactifications \cite{Vafa:1996xn, Denef:2008wq, Grimm:2010ks, Weigand:2010wm}, whose explicit moduli dependence can in general not be obtained beyond the weak coupling regime \cite{Sen:1996vd, Sen:1997gv}. 
This is in contrast to $\mathcal{N}=2$ type IIB/F-theory compactifications where exact in $g_s$ (albeit perturbatively exact in $\alpha'$) results have been found \cite{GarciaEtxebarria:2012zm}. Thus, understanding the possible perturbative corrections for $\mathcal{N}=1$ vacua is certainly a much welcome step for string theory to make contact with particle physics and cosmology.

A powerful tool to understand the effective action of string theory is duality. Recently, a program to derive $\alpha'$ corrections to F-theory compactifications was initiated in \cite{GarciaEtxebarria:2012zm}, where it was demonstrated that such corrections can be obtained through a chain of string/M/F-theory dualities. Subsequently, it was argued in \cite{Grimm:2013gma, Grimm:2013bha} (using techniques of earlier works such as \cite{Grimm:2012rg}) that the K{\"{a}}hler potential of $\mathcal{N}=1$ F-theory compactifications receives a previously unknown correction at order $\alpha^{\prime 2}$ and open string tree-level.
The correction can be derived from a dimensional reduction of higher-derivative curvature corrections in M-theory \cite{Vafa:1995fj, Duff:1995wd, Green:1997di, Green:1997as, Kiritsis:1997em, Russo:1997mk, Antoniadis:1997eg, Tseytlin:2000sf, Liu:2013dna} and induces an Einstein-Hilbert term in the F-theory effective action via the M-theory/F-theory duality. This was interpreted in \cite{Grimm:2013gma} as a shift in the classical compactification volume and, hence, the K{\"{a}}hler potential by a term proportional to the intersection volume of D$7$-branes with an O$7$-plane.
Similar volume corrections were also proposed earlier in M-theory compactifications on Calabi-Yau fourfolds \cite{Haack:2001jz}. The potential consequences for moduli stabilization in type IIB were analyzed in \cite{Pedro:2013qga} under the assumption that the K{\"{a}}hler coordinates are not modified. There, it was argued that KKLT-like minima would survive the correction only at the cost of extreme fine-tuning of the flux superpotential and that strong lower bounds are imposed on the required volumes in the large volume and K{\"{a}}hler uplifting scenarios.
A more careful analysis taking into account additional M-theory terms \cite{Grimm:2013bha} then revealed, however, that the volume shift should be interpreted as a correction to the definition of the K{\"{a}}hler coordinates such that the K{\"{a}}hler potential remains of the classical no-scale form.

Nevertheless, the state of affairs is not fully satisfactory. For one thing, no systematic way was offered to diagnose whether the corrections obtained by dualizing a selected number of terms are genuine or fake (i.e., can be removed by an 11D field redefinition). Moreover, it is not obvious whether additional terms in the 4D scalar potential such as instanton effects couple to the corrected or to the classical K{\"{a}}hler coordinates. In the latter case, the correction of \cite{Grimm:2013gma, Grimm:2013bha} could still have a significant effect on moduli stabilization even though it does not by itself break the classical no-scale structure. Furthermore, the appearance of the correction is quite surprising from the point of view of type IIB string theory. The authors of \cite{Grimm:2013gma, Grimm:2013bha} conjectured that it arises at tree-level in $g_s$ from worldsheets involving open strings, and so it should be apparent already in the weakly coupled type IIB limit. However, a tree-level Einstein-Hilbert term on the worldvolume of a brane only appears in bosonic string theory \cite{Corley:2001hg} but not in superstring theory \cite{Bachas:1999um}. This is true even in the presence of non-trivial brane intersections: while scattering analyses on intersecting D$6$-branes/O$6$-planes show that an Einstein-Hilbert term can be induced at brane intersections at one-loop order \cite{Epple:2004ra}, there does not seem to exist any string diagram that has the right form to generate such a term already at tree-level. If the correction of \cite{Grimm:2013gma, Grimm:2013bha} is real, this needs to be explained, independent of whether it is a correction to the K{\"{a}}hler potential or the K{\"{a}}hler coordinates. An interesting possibility would be a genuinely new F-theory effect that is simply not captured by a perturbative type IIB analysis. That such effects can arise even at weak coupling was discussed in \cite{Collinucci:2008pf}. We argue in this paper, however, that this is not the case here. Instead, the disagreement betweeen the M/F-theory analysis and the type IIB picture is resolved by an 11D field redefinition, as we will explain below.

Apart from that, it should also be important to understand whether other corrections to the volume dependence of the K{\"{a}}hler potential can arise from D-branes and O-planes and whether such corrections are specific to models with D$7$-branes/O$7$-planes or present in more general setups. A crucial question for model building and moduli stabilization is furthermore which degrees of freedom are involved. In particular, the F-theory perspective tends to obscure the open/closed string distinction since the Calabi-Yau fourfold data encodes degrees of freedom of both sectors. For these reasons, we consider it important to try to understand the correction of \cite{Grimm:2013gma, Grimm:2013bha} and other possible volume corrections to the K{\"{a}}hler potential from the point of view of type II string theory.

In the first part of our paper, we will explain that the correction proposed in \cite{Grimm:2013gma, Grimm:2013bha} is an artifact of an inconvenient choice for the basis of the M-theory fields and can be removed by a field redefinition of the metric. We emphasize that this artifact of being led to a misleading frame upon duality transformation from M to F-theory already exists at the 10D/11D level prior to compactification and does not rely on having an effective 4D description. Thus, the diagnoses we present here can be used to systematically distinguish between fake and genuine corrections in a wider context.

Let us clarify here how our result differs from and goes beyond the 4D field redefinition of the K{\"{a}}hler coordinates already found in \cite{Grimm:2013bha}. Although also referred to as a field redefinition, the latter is not optional but a required correction to the definition of the K{\"{a}}hler coordinates that arises from the dimensional reduction of certain M-theory terms in a specific frame. The 11D metric redefinition we find, on the other hand, is a choice one is free to make. In particular, one may choose a convenient frame in which all M-theory terms relevant for the presence of the correction of \cite{Grimm:2013gma, Grimm:2013bha} vanish and neither the 4D K{\"{a}}hler potential nor the K{\"{a}}hler coordinates are corrected. Furthermore, as is well-known in the literature, field redefinitions such as the one we find generate terms in the effective action that are proportional to the leading order equations of motion and thus vanish on-shell. Such terms are not determined by the string S-matrix and are thus arbitrary from the point of view of string scattering. In particular, they are not associated to any string diagram. Our finding thus resolves the tension between the results of \cite{Grimm:2013gma, Grimm:2013bha} and the difficulty in reproducing these results from the point of view of type IIB strings.

We then turn to the question whether there might be further corrections to the volume dependence of the K{\"{a}}hler potential related to the presence of D-branes and O-planes. While we do not present a full answer to this problem, we argue that some corrections can be understood in the type II language as arising from the dimensional reduction of curvature corrections to the Dirac-Born-Infeld (DBI) and Wess-Zumino (WZ) actions. Somewhat surprisingly, the effect of such corrections on the 4D effective action has, to the best of our knowledge, not been explored in the literature so far. The aim of this work is therefore to initiate such a study. Deriving the effective action for general brane and flux configurations in type IIA and IIB is a formidable task already at tree-level (see, e.g., \cite{Grana:2003ek, Camara:2003ku, Camara:2004jj, Lust:2004fi, Jockers:2004yj, Jockers:2005zy, Grimm:2004uq, Grimm:2004ua, Grimm:2008dq, Blumenhagen:2006ci}). In general warped backgrounds, this is further complicated by many additional subtleties (see, e.g., \cite{DeWolfe:2002nn, Giddings:2005ff, Frey:2006wv, Burgess:2006mn, Douglas:2007tu, Koerber:2007xk, Shiu:2008ry, Douglas:2008jx, Frey:2008xw, Marchesano:2008rg, Martucci:2009sf, Chen:2009zi, Douglas:2009zn, Douglas:2010rt, Blaback:2010sj, Underwood:2010pm, Marchesano:2010bs, Blaback:2012mu, Frey:2013bha, Junghans:2014xfa} for related works). We will therefore not attempt to present an exhaustive treatment here but restrict ourselves to discussing some general mechanisms and leave the computation of the corrections in concrete models for future work.
Along the way, we will comment on possible extensions of our work in the perturbative type IIA/IIB framework as well as on generalizations to F-theory.

We identify two mechanisms that can lead to corrections to the volume dependence of the K{\"{a}}hler potential. The first mechanism relies on an induced Einstein-Hilbert term on the worldvolumes of D$p$-branes and O$p$-planes with $p\ge3$. Such a term can arise from curvature corrections to the DBI action if the string frame metric of the background is non-trivially warped due to the presence of other branes or fluxes in the compactification. From the point of view of the 4D effective field theory, this can then lead to a shift of the compactification volume away from its classical value and, hence, a correction to the K{\"{a}}hler potential. A careful counting of the powers of $g_s$ reveals that such corrections arise at the one-loop order, while the order in $\alpha^{\prime}$ depends on the dimensions of the involved branes. In the special case of intersecting D$7$-branes/O$7$-planes, corrections can appear already at order $\alpha^{\prime 2} g_s^2$, where a model-dependent prefactor is not fixed by our general argument.
If these corrections are non-zero in a specific compactification and not removable by a field redefinition, they could thus potentially spoil moduli stabilization. Due to their loop suppression combined with an extended no-scale structure \cite{Berg:2007wt, Cicoli:2007xp}, however, they would then still be subleading in the scalar potential such that dangerous effects of the kind discussed in \cite{Pedro:2013qga} are not expected. The corrections we find are due to the effect on a D-brane or O-plane from a warp factor generated by another D-brane or O-plane and can be interpreted either as a tree-level exchange of gravitons or as an open or closed string one-loop effect.
This is somewhat reminiscent of an analysis done in \cite{Baumann:2006th}, where the authors performed a closed string channel analysis of the backreaction of a D$3$-brane on a nearby D$7$-brane or Euclidean D$3$-brane in a Calabi-Yau background, thereby confirming and generalizing a corresponding open string one-loop computation on toroidal orientifolds in \cite{Berg:2004ek}.

Independently, we also discuss a second mechanism, which relies on an induced D3-brane charge on the worldvolume of D7-branes. It is well-known in the literature that the volume dependence of the K{\"{a}}hler potential is modified in the presence of D3-branes or other objects carrying D3-brane charge such as 3-form flux \cite{Grana:2003ek, DeWolfe:2002nn, Kachru:2003sx}. Here, we point out that such a correction is also expected to arise from curvature corrections to the WZ action of D$7$-branes, which induce a D$3$-brane charge proportional to the Euler characteristic of the wrapped 4-cycle. This leads to a shift in the definition of the K{\"{a}}hler coordinates such that open and closed string degrees of freedom mix in the K{\"{a}}hler potential, while the compactification volume is left invariant. Contrary to the corrections discussed above, this correction appears at open string tree-level.

This paper is organized as follows. In Section \ref{action}, we state our conventions and discuss the effective actions of D-branes and O-planes in type II string theory along with their known higher-derivative curvature corrections. In Section \ref{fieldredef}, we adress the correction to the K{\"{a}}hler coordinates proposed in \cite{Grimm:2013gma, Grimm:2013bha} and show that the relevant terms can be removed by a field redefinition of the metric in M-theory. In Section \ref{eh}, we argue that warping effects can induce an Einstein-Hilbert term in the DBI action of D-branes and O-planes, which can lead to corrections to the K{\"{a}}hler potential at the one-loop order. We then discuss the example of intersecting D$7$-branes/O$7$-planes and argue that corrections can then already appear at order $\alpha^{\prime 2} g_s^2$ but are suppressed in the scalar potential due to an extended no-scale structure. In Section \ref{d3}, we review how the presence of objects carrying D$3$-brane charge in $\mathcal{N}=1$ type IIB compactifications leads to a mixing of bulk K{\"{a}}hler coordinates and open string moduli in the K{\"{a}}hler potential. We then argue that curvature corrections to the WZ action of D$7$-branes are expected to lead to a mixing of the same kind. We conclude in Section \ref{conclusions} with a brief discussion of our results.\\

\section{D-brane and O-plane effective actions}
\label{action}

The action of a stack of $N$ D$p$-branes \cite{Polchinski:1995mt, Witten:1995im, Polchinski:1996na} in string frame is
\begin{align}
S_{\textrm{D}p} &= - \mu_p \int_\mathcal{W} \d^{p+1} \xi \,\e^{-\phi} \,\mathrm{Tr} \sqrt{-\det (P[g_{\mu\nu} + B_{\mu\nu}] + 2\pi\alpha^{\prime} F_{\mu\nu}) } \nl + \mu_p \int_\mathcal{W} \mathrm{Tr} P[C \w \e^{B}] \w \e^{2\pi\alpha^\prime F}, \label{dpaction}
\end{align}
where the first term is the Dirac-Born-Infeld (DBI) part and the second one the Chern-Simons or Wess-Zumino (WZ) part of the action. The brane charge is given by $\mu_p = (2\pi)^{-p}\alpha^{\prime -(p+1)/2}$. Furthermore, $g$ denotes the spacetime metric, $\phi$ the dilaton, $B$ the NSNS 2-form and $F$ the $U(N)$ gauge field strength on the brane worldvolume. $C=\sum_{n} C_{n}$ is a polyform defined as a formal sum over all RR potentials, where $n$ runs over odd (even) numbers in type IIA (type IIB) string theory.
$P[\ldots]$ denotes the pullback of the fields from the 10D spacetime to the brane worldvolume $\mathcal{W}$. Since branes are dynamical objects, the map defining the pullback is not rigid but subject to fluctuations. These are described by $(9-p)$ scalar fields $\Phi^m$ on the worldvolume, which take values in the adjoint representation of the gauge group $U(N)$. The dependence of \eqref{dpaction} on these scalars is implicit in $P[\ldots]$ and can be computed explicitly via a normal coordinate expansion around the background configuration, as explained, e.g., in \cite{Grana:2003ek}. In the above expression, we have neglected interaction terms involving commutators of the non-abelian scalar fields \cite{Myers:1999ps}.

The action of an O$p^-$-plane takes the form\begin{align}
S_{\textrm{O}p} &= 2^{p-4} \mu_p \int_\mathcal{W^\prime} \d^{p+1} \xi \,\e^{-\phi} \sqrt{-\det (P[g_{\mu\nu}]) } - 2^{p-4} \mu_p \int_\mathcal{W^\prime} P[C_{p+1}], \label{opaction}
\end{align}
where $P[\ldots]$ denotes the (rigid) pullback to the O-plane worldvolume $\mathcal{W^\prime}$.\footnote{Note that the O$p$-plane charge equals $2^{p-4}$ times the D$p$-brane charge on the double cover of the orientifold, whereas it equals $2^{p-5}$ times the D$p$-brane charge on the orientifolded space itself.}

Both the DBI and WZ parts of the brane action \eqref{dpaction} receive higher-derivative curvature corrections, which have been derived from supersymmetry and duality arguments, the requirement of anomaly cancellation and by computing scattering amplitudes \cite{Bachas:1999um, Bershadsky:1995qy, Green:1996dd, Cheung:1997az, Minasian:1997mm, Fotopoulos:2001pt, Wyllard:2001ye, Fotopoulos:2002wy, Morales:1998ux, Stefanski:1998he, Craps:1998fn, Craps:1998tw, Scrucca:1999uz}. Furthermore, it follows from similar arguments that also the action of an O-plane must receive curvature corrections \cite{Morales:1998ux, Stefanski:1998he, Craps:1998fn, Craps:1998tw, Scrucca:1999uz, Dasgupta:1997cd, Dasgupta:1997wd, Schnitzer:2002rt}.

For the DBI action of a D$p$-brane, one finds the string frame correction \cite{Bachas:1999um}
\begin{align}
\delta S_{\textrm{D}p,\textrm{DBI}} = \frac{(2\pi)^4\alpha^{\prime 2}}{24\cdot 32\pi^2} \, \mu_p \int_\mathcal{W} \d^{p+1} \xi\, \e^{-\phi} \sqrt{g} &\left[ R_{\alpha\beta\gamma\delta}R^{\alpha\beta\gamma\delta} - 2 R_{\alpha\beta} R^{\alpha\beta} - R_{ab\gamma\delta} R^{ab\gamma\delta} \right. \nll + \left. 2 R_{ab} R^{ab} \right], \label{dbi-correction}
\end{align}
where $\alpha,\beta=0,\ldots,p$ denote indices tangent to the worldvolume and $a,b=p+1,\ldots,9$ are normal indices. The Ricci tensors are contracted using only tangent indices, i.e., $R_{\alpha\beta} = R_{\alpha}\vphantom{}^\gamma\vphantom{}_{\beta\gamma}$ and $R_{ab} = R_{a}\vphantom{}^\gamma\vphantom{}_{b \gamma}$. The Gau\ss-Codazzi equations relate the Riemann tensors constructed from the tangent and normal bundles on the worldvolume to the pullbacks of Riemann tensors constructed from the spacetime metric, as explained in standard textbooks on differential geometry.

Considerations involving supersymmetry and tadpole cancellation suggest that a correction analogous to \eqref{dbi-correction} should be present in the DBI action of an O$p$-plane (see, e.g., discussions in \cite{Bachas:1999um, Giddings:2001yu}).
These lead to the expectation that the curvature correction to the O-plane DBI action equals $2^{p-5}$ times the correction to the D-brane DBI action \eqref{dbi-correction}. Indeed, this assumption was shown to be correct in \cite{Schnitzer:2002rt} by an explicit computation of scattering amplitudes (see also \cite{Garousi:2006zh}) and recently confirmed in \cite{Robbins:2014ara} using T-duality.
Hence,
\begin{align}
\delta S_{\textrm{O}p,\textrm{DBI}} = \frac{(2\pi)^4\alpha^{\prime 2}}{24\cdot 32\pi^2} \, 2^{p-5} \mu_p \int_\mathcal{W^\prime} \d^{p+1} \xi\, \e^{-\phi} \sqrt{g} &\left[ R_{\alpha\beta\gamma\delta}R^{\alpha\beta\gamma\delta} - 2 R_{\alpha\beta} R^{\alpha\beta} - R_{ab\gamma\delta} R^{ab\gamma\delta} \right. \nll + \left. 2 R_{ab} R^{ab} \right]. \label{dbi-correction-o-plane}
\end{align}

In addition to the above corrections to the DBI action, also the WZ action is corrected at order $\alpha^{\prime 2}$. The WZ corrections depend on the curvature 2-forms of the tangent and normal bundles of the D-brane/O-plane worldvolumes and can be written in terms of certain characteristic classes. Some general information about characteristic classes can be found, e.g., in \cite{Denef:2008wq, Nakahara:2003nw}. The full WZ action of a D$p$-brane including curvature corrections takes the form \cite{Cheung:1997az}
\begin{equation}
S_{\textrm{D}p,\textrm{WZ}} = \mu_p \int_\mathcal{W} \mathrm{Tr} P[C \w \e^{B}] \w \e^{2\pi\alpha^\prime F} \w \sqrt{\frac{\widehat A( (2\pi)^2\alpha^{\prime}\, T\mathcal{W})}{\widehat A( (2\pi)^2\alpha^{\prime}\, N\mathcal{W})}},
\end{equation}
where $\widehat A$ is the $A$-roof genus and $T\mathcal{W}$ and $N\mathcal{W}$ denote the tangent and normal bundles of the worldvolume $\mathcal{W}$. The $\alpha^{\prime 2}$ correction can be extracted from this expression by expanding the A-roof genus in terms of Pontryagin classes $p_1(T\mathcal{W})$ and $p_1(N\mathcal{W})$. This yields
\begin{equation}
\delta S_{\textrm{D}p,\textrm{WZ}} = \frac{(2\pi)^4\alpha^{\prime 2}}{48} \mu_p \int_\mathcal{W} \mathrm{Tr} P[C \w \e^{B}] \w \e^{2\pi\alpha^\prime F} \w \left( p_1(N\mathcal{W}) - p_1(T\mathcal{W}) \right). \label{aa}
\end{equation}
For an O$p^-$-plane, the curvature-corrected WZ action reads
\begin{equation}
S_{\textrm{O}p,\textrm{WZ}} = -2^{p-4} \mu_p \int_\mathcal{W^\prime} P[C] \w \sqrt{\frac{L(\frac{1}{4} (2\pi)^2\alpha^{\prime}\, T\mathcal{W^\prime})}{L(\frac{1}{4} (2\pi)^2\alpha^{\prime}\, N\mathcal{W^\prime})}},
\end{equation}
where $L$ is the Hirzebruch $L$-polynomial. Again expanding the square root in terms of Pontryagin classes and picking out the $\alpha^{\prime 2}$ correction yields
\begin{equation}
\delta S_{\textrm{O}p,\textrm{WZ}} = 2^{p-5} \frac{(2\pi)^4\alpha^{\prime 2}}{48} \mu_p \int_\mathcal{W^\prime} \mathrm{Tr} P[C] \w \left( p_1(N\mathcal{W^\prime}) - p_1(T\mathcal{W^\prime}) \right). \label{cc}
\end{equation}

It is important to stress that neither the D-brane nor the O-plane corrections stated above are complete at order $\alpha^{\prime 2}$.
One indication of this is that neither of the above corrections is properly T-duality covariantized, as was first noted in \cite{Myers:1999ps}. According to the Buscher rules \cite{Buscher:1987sk}, T-duality mixes metric degrees of freedom with the dilaton and the $B$ field. One therefore expects that promoting the above corrections to T-duality invariants produces additional terms involving derivatives of the dilaton and the $B$ field. Due to the requirement of gauge invariance, one then also expects further terms involving derivatives of the worldvolume gauge field strength. In addition, other T-duality and gauge invariant combinations of fields may appear at the same and at higher orders in $\alpha^\prime$. Some of these corrections were computed in \cite{Robbins:2014ara, Wyllard:2000qe, Wijnholt:2003pw, Becker:2010ij, Becker:2011ar, Garousi:2009dj, Garousi:2010ki, Garousi:2010rn,  Garousi:2010bm, Garousi:2011ut, Garousi:2011fc, Velni:2012sv, Velni:2013jha, Hatefi:2010ik, Hatefi:2012ve, Hatefi:2012zh}. Furthermore, the curvature correction to the DBI action may contain additional terms involving the second fundamental form \cite{Bachas:1999um}. Some but not all of these extra terms were shown to vanish in \cite{Fotopoulos:2001pt, Wyllard:2001ye}. Finally, an open issue is to promote all known higher order couplings to a fully non-abelian action as in \cite{Myers:1999ps}.

\section{Coordinate artifacts}
\label{fieldredef}

Let us now discuss the correction to the K{\"{a}}hler coordinates that was argued in \cite{Grimm:2013gma, Grimm:2013bha} to arise at order $\alpha^{\prime 2}g_s$ from D7-brane/O7-plane intersections in the 4D effective action of $\mathcal{N}=1$ F-theory compactifications. Its presence was inferred by the authors by dimensionally reducing a number of curvature corrections to the M-theory action and then exploiting the duality of F-theory with M-theory compactified on a Calabi-Yau 4-fold. As stated in the introduction, the correction can be removed by performing a field redefinition of the metric on the M-theory side of the duality. From the point of view of string perturbation theory, it is therefore arbitrary and not associated to any string diagram (see, e.g., discussions in \cite{Liu:2013dna, Gross:1986iv, Tseytlin:1986zz, Tseytlin:1993df, Forger:1996vj}). This can be seen as follows. Consider a Lagrangian $\mathcal{L}(\phi)$ for a field $\phi$ and perform a field redefinition $\phi \to \phi+\delta\phi$, where $\delta\phi$ is assumed to be small. Expanding the redefined Lagrangian about the original one then yields
\begin{align}
\mathcal{L}(\phi+\delta\phi) &= \mathcal{L}(\phi) + \frac{\delta \mathcal{L}(\phi)}{\delta \phi} \delta \phi + \mathcal{O}\left(\left(\delta\phi\right)^2\right).
\end{align}
If we are only interested in the effective action up to linear order in $\delta \phi$, the field redefinition thus amounts to adding a term to the action that is proportional to the leading order equations of motion. Such terms vanish on-shell and are therefore not fixed by computing string S-matrix elements. In the absence of an off-shell formalism such as string field theory that could fix such terms, they are arbitrary and can be discarded.\footnote{Note that the possibility to define away a term in the M-theory action implies the possibility to define it away in string theory since our field redefinition is valid for any volume of the M-theory circle.}

In order to show that this is true for the correction of \cite{Grimm:2013gma, Grimm:2013bha}, we consider the M-theory action together with the higher-derivative curvature corrections relevant for our discussion \cite{Vafa:1995fj, Duff:1995wd, Green:1997di, Green:1997as, Kiritsis:1997em, Russo:1997mk, Antoniadis:1997eg, Tseytlin:2000sf, Liu:2013dna}. Adopting the conventions of \cite{Grimm:2013bha}, the action reads
\begin{align}
S_{11} = & \frac{1}{2\kappa_{11}^2} \int \d^{11}x \sqrt{g_{11}} \left[ R + k_1 \left( t_8t_8 R^4 - \frac{1}{4!} \epsilon_{11}\epsilon_{11} R^4 \right) \right] \notag \\
& - \frac{1}{2} \frac{1}{2\kappa_{11}^2} \int \d^{11}x \sqrt{g_{11}} \left[ |G_4|^2 + 2k_1 \left( t_8t_8 G_4^2 R^3 + \frac{1}{96} \epsilon_{11}\epsilon_{11} G_4^2 R^3 \right) \right] + \ldots, \label{mtheoryaction}
\end{align}
where $R$ denotes the Ricci scalar of the 11D spacetime, $G_4$ is the 4-form field strength and $\kappa_{11}^2$ and $k_1$ are constants. The higher-derivative corrections consist of various contractions of Riemann tensors and powers of $G_4$ and are stated in the above equation in a shortcut notation involving the 11D epsilon tensor $\epsilon_{11}$ and the so-called $t_8$ tensor. We refer to \cite{Grimm:2013bha} for a detailed explanation of this notation.

Following \cite{Grimm:2013gma, Grimm:2013bha}, we now specialize to the case relevant for the duality with F-theory, where the 11D spacetime is a direct product of an external 3D spacetime and a Calabi-Yau 4-fold. The Ricci scalar can then be split into an external and an internal piece,
\begin{equation}
R = R^{(3)} + R^{(8)},
\end{equation}
where the latter vanishes on-shell by the Calabi-Yau condition. $G_4$ can be decomposed into a purely internal term and a sum of terms with mixed legs,
\begin{equation}
G_4 = G^{(8)}_4 + \sum_i F^{(3) i}_2 \w \omega^{(8) i}_2,
\end{equation}
where the latter can each be written as a product of an external 2-form and an internal $(1,1)$-form. The mixed components of $G_4$ thus have the index structure $G_{\mu\nu \bar m n}$. Since the internal part of $G_4$ was not relevant for the computation in \cite{Grimm:2013gma, Grimm:2013bha}, we will neglect it in the following and assume that $G_4$ only has mixed components.\footnote{Accordingly, we do not consider possible corrections arising from non-trivial 4-form flux (e.g., warping corrections), as they were not considered in \cite{Grimm:2013gma, Grimm:2013bha} either and are therefore not relevant for the present purpose.}

Writing out the higher-derivative terms in \eqref{mtheoryaction} explicitly yields a huge number of different contractions of Riemann tensors and $G_4$ field strengths. However, as derived in detail in \cite{Grimm:2013gma, Grimm:2013bha}, the correction to the K{\"{a}}hler coordinates in the F-theory effective action descends from only two types of terms in \eqref{mtheoryaction}. The first type consists of one external Ricci scalar multiplied by a contraction of 3 internal Riemann tensors, and the second type is a contraction of two mixed $G_4$ factors with 3 internal Riemann tensors.
Using a computer algebra system such as Cadabra \cite{Peeters:2007wn, Peeters2}, it is straightforward to isolate the relevant terms in the M-theory action. One finds
\begin{align}
& k_1 \left( t_8t_8 R^4 - \frac{1}{4!} \epsilon_{11}\epsilon_{11} R^4 \right) \supset 256 k_1 R^{(3)} \left( R^n\vphantom{}_{mp}\vphantom{}^q R^m\vphantom{}_{nr}\vphantom{}^s R^p\vphantom{}_{qs}\vphantom{}^r + R^n\vphantom{}_{mp}\vphantom{}^q R^r\vphantom{}_{nq}\vphantom{}^s R^m\vphantom{}_{rs}\vphantom{}^p \right)
, \label{corr1} \\
& 2 k_1 \left( t_8t_8 G_4^2 R^3 + \frac{1}{96} \epsilon_{11}\epsilon_{11} G_4^2 R^3 \right) \supset -256 k_1 |G_4|^2 \left( R^n\vphantom{}_{mp}\vphantom{}^q R^m\vphantom{}_{nr}\vphantom{}^s R^p\vphantom{}_{qs}\vphantom{}^r + R^n\vphantom{}_{mp}\vphantom{}^q R^r\vphantom{}_{nq}\vphantom{}^s R^m\vphantom{}_{rs}\vphantom{}^p \right) \nl
- 6 \cdot 256 k_1 |G_4|^2_{m \bar l} \,g^{\bar ln} \left( R^m\vphantom{}_{np}\vphantom{}^q R^r\vphantom{}_{qs}\vphantom{}^t R^p\vphantom{}_{rt}\vphantom{}^s - R^p\vphantom{}_{nq}\vphantom{}^r R^m\vphantom{}_{ps}\vphantom{}^t R^q\vphantom{}_{rt}\vphantom{}^s - R^p\vphantom{}_{nq}\vphantom{}^r R^s\vphantom{}_{pr}\vphantom{}^t R^m\vphantom{}_{st}\vphantom{}^q \right), \label{corr2}
\end{align}
where we have written the internal terms in complex coordinates and used the symmetries of the Riemann tensor to simplify the expressions. All other terms are either not of the desired form or contain powers of internal Ricci tensors, which vanish on-shell by the Calabi-Yau condition and can therefore be neglected.

As stated above, we can remove the terms on the right-hand sides of \eqref{corr1} and \eqref{corr2} from the action \eqref{mtheoryaction} by performing a field redefinition of the form
\begin{equation}
g_{\bar m n} \to g_{\bar m n} + h_{\bar m n}, \label{redef}
\end{equation}
where $h_{\bar m n}$ is a sum of contractions of 3 internal Riemann tensors. Expanding \eqref{mtheoryaction} to linear order in $h$, we find that the redefinition \eqref{redef} modifies the leading order Einstein-Hilbert and $|G_4|^2$ terms in the action according to
\begin{align}
\sqrt{g_{11}}\, R^{(3)} &\to \sqrt{g_{11}} \left[ R^{(3)} + \frac{1}{2}\left(h^m_m + h^{\bar m}_{\bar m}\right) R^{(3)} \right], \label{redef1} \\
\sqrt{g_{11}}\, |G_4|^2 &\to \sqrt{g_{11}} \left[ |G_4|^2 + \frac{1}{2}\left(h^m_m + h^{\bar m}_{\bar m}\right) |G_4|^2 - 2|G_4|^2_{m \bar l} h^{m \bar l} \right] \label{redef2}
\end{align}
up to terms that would be of higher order in derivatives of the fields and can therefore be discarded.

Comparing \eqref{redef1} and \eqref{redef2} with the right-hand sides of \eqref{corr1} and \eqref{corr2}, one verifies that the latter can be removed from the action by choosing
\begin{equation}
h_{\bar m n} = H g_{\bar m n} + K_{\bar m n},
\end{equation}
where
\begin{align}
H &= 256 k_1 \left( R^n\vphantom{}_{mp}\vphantom{}^q R^m\vphantom{}_{nr}\vphantom{}^s R^p\vphantom{}_{qs}\vphantom{}^r + R^n\vphantom{}_{mp}\vphantom{}^q R^r\vphantom{}_{nq}\vphantom{}^s R^m\vphantom{}_{rs}\vphantom{}^p \right), \\
K_{\bar m n} &= 768 k_1 \,g_{\bar m l} \left( R^l\vphantom{}_{np}\vphantom{}^q R^r\vphantom{}_{qs}\vphantom{}^t R^p\vphantom{}_{rt}\vphantom{}^s - R^p\vphantom{}_{nq}\vphantom{}^r R^l\vphantom{}_{ps}\vphantom{}^t R^q\vphantom{}_{rt}\vphantom{}^s - R^p\vphantom{}_{nq}\vphantom{}^r R^s\vphantom{}_{pr}\vphantom{}^t R^l\vphantom{}_{st}\vphantom{}^q \right).
\end{align}
Hence, we have shown that the correction to the F-theory effective action proposed in \cite{Grimm:2013gma, Grimm:2013bha} arises from an inconvenient choice of the fields on the dual M-theory side. Switching to a more convenient basis, it can be discarded and does not appear in the effective action anymore.

Let us stress again that this result goes beyond the earlier result of \cite{Grimm:2013bha}, where it was shown that the correction found in \cite{Grimm:2013gma} amounts to a redefinition of the 4D K{\"{a}}hler coordinates. Corrections to the K{\"{a}}hler coordinates do in general not have to be removable by a field redefinition in the 10D/11D parent theory. A well-known counter-example are the corrections due to the presence of D3-branes, which involve a shift in the definition of the K{\"{a}}hler coordinates that mixes open and closed string degrees of freedom \cite{Grana:2003ek, DeWolfe:2002nn, Kachru:2003sx}. These shifts cannot simply be removed by going to a different 10D frame but are important for the consistency of the theory (cf. Section \ref{d3}). A physical effect arising from the shifts is, e.g., the well-known $\eta$ problem in warped brane inflation \cite{Kachru:2003sx}. Hence, the results of \cite{Grimm:2013bha} do a priori not imply that the correction is an artifact of an inconvenient 10D/11D field basis. On the contrary, it was proposed in \cite{Grimm:2013gma, Grimm:2013bha} that the correction is due to open string diagrams, which is incompatible with the statement that it can be removed by a 10D/11D field redefinition. The computation in this section shows that such a field redefinition exists and thus reconciles the M/F-theory computation of \cite{Grimm:2013gma, Grimm:2013bha} with the perturbative type IIB picture.

We finally comment on the fact that we do not consider warping in this section. As discussed above, the main question we wanted to address here is whether the correction found in \cite{Grimm:2013gma, Grimm:2013bha} is due to open strings stretching between intersecting 7-branes. The answer to this question should be independent of whether a given compactification is warped or not since 7-branes do by themselves not induce any warping (unless they carry a D3-brane charge, which is model-dependent). Instead, warping is induced by, e.g., D3-branes, O3-planes and 3-form flux, and its presence and strength depends on how these objects are distributed in a given compactification. Possible corrections due to 7-brane intersections should therefore already be seen in the unwarped case and, hence, for simplicity, we can focus on a setting where warping is absent or weak. This is consistent with the fact that warping was not considered in \cite{Grimm:2013gma, Grimm:2013bha} either, which shows that their correction is clearly not a warping effect. We stress, however, that, in a highly warped setting, the K{\"{a}}hler potential can receive large corrections on top of the one of \cite{Grimm:2013gma, Grimm:2013bha}, and one would have to redo the analysis of \cite{Grimm:2013gma, Grimm:2013bha} as well as the computation in this section to find its full form. Such a computation is beyond the scope of this work. For the above reasons, it would be surprising if our overall conclusions were affected by warping although the potential is likely to be more complicated.

\section[{K{\"{a}}hler corrections from induced Einstein-Hilbert terms}]{K{\"{a}}hler corrections from induced Einstein-Hilbert\\terms}
\label{eh}

{ In view of the proposal of \cite{Grimm:2013gma, Grimm:2013bha}, one might wonder whether type II string theory yields other corrections to the K{\"{a}}hler potential that are of a similar form, i.e., arise on intersections of D-branes and O-planes and correct the classical compactification volume. In this section, we argue that curvature corrections to the DBI action of spacetime-filling D$p$-branes or O$p$-planes with $p\ge3$ can induce an Einstein-Hilbert term in the 4D effective action of type II string compactifications. This would imply a shift in the classical compactification volume and is therefore a possible mechanism to generate corrections to the K{\"{a}}hler potential. The effect can arise from $R^2$ curvature corrections \cite{Bachas:1999um} if the string frame metric is non-trivially warped, where its presence and magnitude is expected to be model-dependent. As warping occurs under rather generic circumstances in compactifications with branes and fluxes (see, e.g., \cite{Giddings:2001yu, Becker:1996gj, Dasgupta:1999ss, Gukov:1999ya, Greene:2000gh}), our discussion applies to a wide range of compactifications in type IIA and type IIB string theory.
We argue that possible corrections of this type, if not removable by a field redefinition, appear earliest at one-loop order and would therefore be suppressed in the K{\"{a}}hler potential compared to the tree-level corrections proposed in \cite{Grimm:2013gma, Grimm:2013bha}. We discuss in detail the example of intersecting D7-branes/O7-planes, where a correction could already arise at order $\alpha^{\prime 2}g_s^2$ and might thus be dangerous for the usual moduli stabilization scenarios. Owing to the presence of an extended no-scale structure \cite{Berg:2007wt, Cicoli:2007xp}, however, we conclude that even in this case the scalar potential would only receive subleading corrections and is therefore not expected to be destabilized.}

\subsection{Induced Einstein-Hilbert term due to warping}

In the presence of fluxes and localized sources such as D-branes and O-planes, string compactifications are generically warped. Let us therefore assume that the string frame metric has the form of a warped product space,
\begin{equation}
\d s_{10}^2 = g_{\mu\nu} \d x^\mu \d x^\nu + g_{mn} \d y^m \d y^n = \e^{2A(y)} \tilde g_{\mu\nu} \d x^\mu \d x^\nu + \e^{2B(y)} \tilde g_{mn} \d y^m \d y^n, \label{warpedmetric}
\end{equation}
where $\mu,\nu =0,\ldots,3$ are external indices and $m,n=4,\ldots,9$ are internal indices. Furthermore, $A$ is the warp factor, which depends non-trivially on the internal coordinates $y^m$, and $B$ is a conformal factor that may be pulled out of the internal metric for convenience. In the special case where $A=B=(\phi-\phi_0)/4$ (with $\phi_0 = \ln g_s$), the tilded metric coincides with the metric in 10D Einstein frame, which, accordingly, is then unwarped. More generally, the string frame warp factor $A$ is related to the warp factor $A_E$ in 10D Einstein frame via $A = A_E + (\phi-\phi_0)/4$. Hence, for a generic warped string frame metric, also the Einstein frame metric is warped and vice versa.

The components of the Riemann tensor for a metric of the form \eqref{warpedmetric} can be computed using standard methods from textbooks such as \cite{Wald:1984rg}. One finds
\begin{align}
R_{\mu\nu\lambda\rho} &= - 2\e^{4A-2B} \tilde g_{\lambda[\mu} \tilde g_{\nu]\rho} (\partial_m A) (\tilde \partial^m A) + \e^{2A} \tilde R_{\mu\nu\lambda\rho}, \label{riemann1} \\
R_{\mu j \lambda l} &= - \e^{2A} \tilde g_{\mu\lambda} \tilde \nabla_j \partial_l A - \e^{2A} \tilde g_{\mu\lambda} (\partial_j A) (\partial_l A) + \e^{2A} \tilde g_{\mu\lambda} (\partial_j B) (\partial_l A) \nl + \e^{2A} \tilde g_{\mu\lambda} (\partial_j A) (\partial_l B) - \e^{2A} \tilde g_{jl} \tilde g_{\mu\lambda} (\partial_m A) (\tilde \partial^m B), \\
R_{ijkl} &= 2 \e^{2B} \tilde g_{l[i} \tilde \nabla_{j]} \partial_k B - 2 \e^{2B} \tilde g_{k[i} \tilde \nabla_{j]} \partial_l B  - 2 \e^{2B} \tilde g_{l[i} (\partial_{j]} B) (\partial_k B) \nl + 2 \e^{2B} \tilde g_{k[i} (\partial_{j]} B) (\partial_l B) - 2 \e^{2B} \tilde g_{k[i} \tilde g_{lj]} (\partial_m B) (\tilde \partial^m B) + \e^{2B} \tilde R_{ijkl}, \label{riemann3}
\end{align}
where all other components are zero (up to components related to the above by symmetries of the Riemann tensor). In the above expressions, indices in brackets are anti-symmetrized with weight one, and tildes on curvature tensors, covariant derivatives and contractions indicate that the objects are constructed using the unwarped metric $\tilde g$. Analogously, indices on tilded objects are raised and lowered with $\tilde g$, while indices on objects without tildes are raised and lowered with $g$.

Let us now consider the curvature correction to the DBI action \eqref{dbi-correction} for a spacetime-filling D$p$-brane in static gauge in a warped background of the form \eqref{warpedmetric}.
The external and internal components of the worldvolume Riemann tensor $R_{\alpha\beta\gamma\delta}$ are then given by expressions similar to \eqref{riemann1}--\eqref{riemann3}, except that the internal indices do not run from $4$ to $9$ but only from $4$ to $p$. We may therefore express $R_{\alpha\beta\gamma\delta}$ in terms of the components of the spacetime Riemann tensor as follows: $R_{\alpha\beta\gamma\delta}$ is obtained by projecting \eqref{riemann1}--\eqref{riemann3} to the worldvolume directions and then subtracting those terms that involve connections to the normal bundle. Hence,
\begin{align}
R_{\alpha\beta\gamma\delta} &= \delta^\mu_\alpha \delta^\nu_\beta \delta^\lambda_\gamma \delta^\rho_\delta\, (R_T)_{\mu\nu\lambda\rho} + 4\delta^{[\mu}_\alpha \delta^{j]}_\beta \delta^{[\lambda}_\gamma \delta^{l]}_\delta\, (R_T)_{\mu j \lambda l} + \delta^i_\alpha \delta^j_\beta \delta^k_\gamma \delta^l_\delta\, (R_T)_{ijkl}, \label{projection}
\end{align}
where we introduced the shortcut notation
\begin{align}
(R_T)_{\mu\nu\lambda\rho} &= R_{\mu\nu\lambda\rho} + 2\e^{4A-2B} \tilde g_{\lambda[\mu} \tilde g_{\nu]\rho} (\tilde{\partial_N A})^2, \label{proj1}\\
(R_T)_{\mu j \lambda l} &= R_{\mu j \lambda l} + \e^{2A} \tilde g_{jl} \tilde g_{\mu\lambda} (\tilde{\partial_N A}) \cdot (\tilde{\partial_N B}), \\
(R_T)_{ijkl} &= R_{ijkl} + 2 \e^{2B} \tilde g_{k[i} \tilde g_{lj]} (\tilde{\partial_N B})^2. \label{proj3}
\end{align}
Here and in the following, $R_T$ denotes the curvature of the tangent bundle, and the notation $\partial_{N (T)}$ indicates that derivatives are taken along directions normal (tangent) to the brane worldvolume. Similar expressions can be derived in order to relate the worldvolume curvature of the normal bundle to the spacetime curvature, but we will not need them in the following.

Since we are interested in how warping induces a 4D Einstein-Hilbert term on the brane worldvolume, let us now analyze those terms in \eqref{dbi-correction} that are contracted with at least one external Riemann tensor $R_{\mu\nu\lambda\rho}$. Substituting \eqref{projection} into \eqref{dbi-correction}, we find
\begin{align}
\delta S_\textrm{DBI} \supset \frac{(2\pi)^4\alpha^{\prime 2}}{24\cdot 32\pi^2}\, \mu_p \int_\mathcal{W} \d^4 x\, \d^{p-3}y\, \e^{-\phi} \sqrt{g} &\left[ (R_T)_{\mu\nu\rho\lambda}(R_T)^{\mu\nu\rho\lambda} - 2 (R_T)_{\mu}\vphantom{}^\rho\vphantom{}_{\nu\rho} (R_T)^{\mu\lambda\nu}\vphantom{}_{\lambda} \right. \nll - \left. 4 (R_T)_{\mu}\vphantom{}^k\vphantom{}_{\nu k} (R_T)^{\mu \lambda\nu}\vphantom{}_{\lambda}\right] \label{warped0}
\end{align}
up to irrelevant terms, where $\mu,\nu=0,\ldots,3$ are external indices and $k,l=4,\ldots,p$ are internal indices along the brane. The metric determinant can be written in terms of the unwarped metric,
\begin{equation}
\sqrt{g} = \sqrt{\tilde g}\, \e^{4A+(p-3)B}. \label{warped2}
\end{equation}
Furthermore, using \eqref{riemann1}--\eqref{riemann3} and \eqref{proj1}--\eqref{proj3}, the terms in the bracket on the right-hand side of \eqref{warped0} can be rewritten as
\begin{align}
\left[ \ldots \right] &= \e^{-4A} \tilde R_{\mu\nu\rho\lambda} \tilde R^{\mu\nu\rho\lambda} - 2 \e^{-4A} \tilde R^{(4)}_{\mu\nu} \tilde R^{(4)\mu\nu} + 12 \e^{-2A-2B} \tilde R^{(4)} (\tilde{\partial_T A})^2 \nl + 4 \e^{-2A-2B} \tilde R^{(4)} \tilde \nabla^2_T A + 4(p-5) \e^{-2A-2B} \tilde R^{(4)} (\tilde{\partial_T A})\cdot (\tilde{\partial_T B}) + \textrm{internal}, \label{warped1}
\end{align}
where $\tilde R^{(4)}$ and $\tilde R^{(4)}_{\mu\nu}$ denote the external Ricci scalar and the external Ricci tensor, respectively.

Substituting \eqref{warped1} and \eqref{warped2} into \eqref{warped0} and picking out the terms linear in $\tilde R^{(4)}$, we find the induced Einstein-Hilbert term
\begin{align}
\Delta S_\textrm{EH} = \frac{(2\pi)^4\alpha^{\prime 2}}{24\cdot 32\pi^2}\, \mu_p \int_\mathcal{W} &\d^4 x\, \d^{p-3}y\, \sqrt{\tilde g}\, \tilde R^{(4)}\, \e^{-\phi+2A+(p-5)B} \nll \times \left[ 12 (\tilde{\partial_T A})^2 + 4 \tilde \nabla^2_T A + 4(p-5) (\tilde{\partial_T A})\cdot (\tilde{\partial_T B}) \right]. \label{induced-eh}
\end{align}
Splitting the metric into a 4D external and a $(p-3)$d internal part, we can write this in the form
\begin{equation}
\Delta S_\textrm{EH} = \frac{1}{2\kappa^2g_s^2} \int \d^4 x \sqrt{\tilde g^{(4)}}\, \Delta \mathcal{V} \tilde R^{(4)}, \label{induced-eh2}
\end{equation}
where we define the volume shift
\begin{align}
\Delta \mathcal{V} = \frac{(2\pi)^4\alpha^{\prime 2}}{24\cdot 32\pi^2}\, 2\kappa^2 g_s^2\, \mu_p \int &\d^{p-3}y\, \sqrt{\tilde g^{(p-3)}}\, \e^{-\phi+2A+(p-5)B}  \nll \times \left[ 12 (\tilde{\partial_T A})^2 + 4 \tilde \nabla^2_T A + 4(p-5) (\tilde{\partial_T A})\cdot (\tilde{\partial_T B}) \right]. \label{induced-eh3}
\end{align}
We have thus shown that non-trivial warping can induce an effective Einstein-Hilbert term on the worldvolume of a D-brane.
Note that an analogous term is induced on the worldvolume of an O-plane, as can easily be seen by repeating the above procedure for the correction \eqref{dbi-correction-o-plane}. 

We also note that the dimensional reduction of the curvature corrections \eqref{dbi-correction} and \eqref{dbi-correction-o-plane} yields a number of further corrections to the 4D effective action, some of which already appear at tree-level. One of them is a term quadratic in the external curvature $\tilde R_{\mu\nu\rho\lambda}$, as follows from substituting \eqref{warped1} into \eqref{warped0}. Specializing to the worldvolume of a D7-brane or O7-plane, we thus reproduce a correction obtained in \cite{Grimm:2013gma}. In addition, the dimensional reduction of \eqref{dbi-correction} and \eqref{dbi-correction-o-plane} corrects the 4D effective action by a variety of purely internal terms which we do not discuss in the following. However, we note that these terms are interesting in their own right as they contribute to the 4D scalar potential.

\subsection{K{\"{a}}hler potential}

Let us now discuss how an induced Einstein-Hilbert term of the form \eqref{induced-eh2} affects the K{\"{a}}hler potential in the 4D effective action. We first consider the Einstein-Hilbert term of the leading order 10D bulk action, which in the string frame reads
\begin{equation}
S_{10} = \frac{1}{2\kappa^2} \int \sqrt{g}\, \e^{-2\phi} R + \ldots \supset \frac{1}{2\kappa^2} \int \sqrt{\tilde g}\, \e^{-2\phi+2A+6B} \tilde R^{(4)},
\end{equation}
where we used \eqref{riemann1}--\eqref{riemann3} in order to pick out the term involving the external Ricci scalar. In order to compare this expression with the existing literature, let us now switch to 10D Einstein frame with $A = A_E + (\phi-\phi_0)/4$, where $A_E$ is the Einstein frame warp factor and $\phi_0 = \ln g_s$. The conformal factor $B$ is just a gauge choice and can be set to $B=(\phi-\phi_0)/4$ for convenience. This yields
\begin{equation}
S_{10} \supset \frac{1}{2\kappa_{10}^2} \int \sqrt{\tilde g^{(4)}}\, \mathcal{V}_\textrm{w} \tilde R^{(4)}, \qquad \mathcal{V}_\textrm{w} = \int \sqrt{\tilde g^{(6)}}\, \e^{2A_E}, \label{bulk-eh}
\end{equation}
where $2\kappa_{10}^2 = 2\kappa^2 g_s^2$ and $\mathcal{V}_\textrm{w}$ is the warped volume of the compactification manifold. The total Einstein-Hilbert term in the 4D effective action including the previously derived correction \eqref{induced-eh2} is therefore
\begin{equation}
S_\textrm{EH} = \frac{1}{2\kappa_{10}^2} \int \sqrt{\tilde g^{(4)}}\, \left( \mathcal{V}_\textrm{w} + \Delta \mathcal{V} \right) \tilde R^{(4)}. \label{bulk-eh2}
\end{equation}

From this expression, we can read off the K{\"{a}}hler potential. In the absence of warping, $\mathcal{V}_\textrm{w}$ reduces to the standard unwarped volume $\mathcal{V} = \int \sqrt{\tilde g^{(6)}}$. The K{\"{a}}hler potential at tree-level is then known to have the usual no-scale form (see, e.g., \cite{Grana:2003ek})
\begin{equation}
K = -2 \ln (\mathcal{V}) + \ldots,
\end{equation}
where the dots indicate contributions from other sectors such as the complex structure moduli.
According to \eqref{bulk-eh2}, however, the presence of warping leads to a shift in the coefficient of the Einstein-Hilbert term,
\begin{equation}
\mathcal{V} \to \mathcal{V}_\textrm{w}+\Delta \mathcal{V}.
\end{equation}
Interpreting this as a shift in the classical volume $\mathcal{V}$, the
K{\"{a}}hler potential is modified according to
\begin{equation}
K = -2 \ln (\mathcal{V}_\textrm{w}+\Delta \mathcal{V}) + \ldots \label{kaehler}
\end{equation}
The first term in the argument of the logarithm is well-known to appear in the presence of (Einstein frame) warping and was discussed, e.g., in \cite{DeWolfe:2002nn, Giddings:2005ff}. The second correction is new and is induced by D-branes and O-planes that are located in warped regions of the compact space. We stress that the correction appears if the \emph{string frame} metric is non-trivially warped and can therefore be non-zero even if the Einstein frame metric is unwarped, i.e., for $A_E=0$.

As the mechanism discussed above is general and not restricted to a specific compactification, we do not evaluate the volume shift \eqref{induced-eh3} explicitly. However, using general arguments, we can infer the parametric form of possible corrections and, in particular, the order in $\alpha^\prime$ and $g_s$ at which they would appear in the 4D effective action. In the following, we will focus on the situation where the warp factor that generates a correction on a D$p$-brane or O$p$-plane is sourced by another D$p^\prime$-brane or O$p^\prime$-plane (where $p^\prime$ can in general differ from $p$). As mentioned above, however, warping can more generally also be sourced by other objects such as RR and NSNS fluxes. Assuming the source to be a D-brane or O-plane, the tree-level bulk equations of motion relate it to the warp factor via
\begin{equation}
\tilde \nabla^2 A \sim 2\kappa^2 g_s\, \mu_{p^\prime}\, \delta^{(9-{p^\prime})},
\end{equation}
where a relative factor of $2\kappa^2 g_s$ appears between the two terms because the left-hand side is derived from a closed string tree-level diagram, whereas the right-hand side comes from tree-level worldsheets with either a boundary or a cross-cap. Hence, we conclude that the on-shell expression for $\tilde \nabla^2 A$ is of the order $2\kappa^2 \mu_{p^\prime} g_s$. Similar remarks apply to other two-derivative terms of the warp factor. Using this together with $2\kappa^2 = (2\pi)^7 (\alpha^\prime)^4$, $\mu_p = (2\pi)^{-p}(\alpha^\prime)^{-(p+1)/2}$ and $\e^{\phi_0} = g_s$ in \eqref{induced-eh3}, we find
\begin{equation}
\Delta \mathcal{V} \sim g_s^2\, \alpha^{\prime 10-(p+1)/2-(p^\prime+1)/2}. \label{dsgusgh}
\end{equation}
Naively, one might have concluded that the volume correction arises already at order $g_s$ since it is derived from the D-brane or O-plane action. However, since the warp factor term is sourced only at order $g_s$ by another D-brane or O-plane, the correction arises at order $g_s^2$ and is therefore a one-loop correction. The power of $\alpha^\prime$ depends on the dimensions of the D-branes or O-planes that source the warping and of those on which the correction is induced. In order to avoid confusion, let us emphasize that, for any $p$ and $p^\prime$, \eqref{dsgusgh} always comes from order $\alpha^{\prime 2}$ curvature corrections to the DBI action. However, in the 4D effective action, the order in $\alpha^\prime$ at which the corrections appear depends on the dimensions of the D-branes and O-planes involved.

As explained above, \eqref{dsgusgh} arises from the effect that the change in the warp factor caused by a localized source has on another localized source. Another way of saying this is that we compute a tree-level exchange of gravitons betweeen two D-branes or O-planes. This can be interpreted as the closed string channel dual of an open or closed string one-loop effect arising from annulus (D-brane/D-brane), M{\"{o}}bius strip (D-brane/O-plane) or Klein bottle diagrams (O-plane/O-plane) (see, e.g., \cite{Kakushadze:2001bd} for a similar discussion). Taking, for example, the case where the correction arises at the intersection of two D-branes, it can either be viewed as being due to a tree-level exchange of gravitons or as a one-loop process of open strings stretching between the branes. This duality is what allows us to infer the presence of a one-loop correction by analyzing tree-level equations of motion of closed string fields. Our method is reminiscent of a similar computation done in \cite{Baumann:2006th}, where the authors computed the perturbation of the warp factor by a D$3$-brane in a non-compact Calabi-Yau background and the effect of this perturbation on the volume of the 4-cycle wrapped by a nearby D$7$-brane or Euclidean D$3$-brane. This was then interpreted as a closed string channel computation of an open string one-loop process computed in \cite{Berg:2004ek}.

A crucial question is obviously whether the general mechanism discussed here is actually realized in concrete string compactifications.
In particular, it is important to check whether corrections of the above form are removable by a field redefinition, as emphasized in Section \ref{fieldredef}.
Furthermore, the supergravity fields appearing in \eqref{induced-eh3} are not fixed by our general arguments, and so one might wonder whether the correction vanishes under certain conditions, e.g., due to a cancellation of terms of opposite signs arising on different branes or due to backreaction effects that let the warped volume factor in \eqref{dsgusgh} go to zero.
In general, we expect this to be a model-dependent question, and we leave a detailed analysis of specific string compactifications for future work.
We emphasize, however, that explicit CFT computations show that an Einstein-Hilbert term can indeed be induced on brane intersections at one-loop order. In \cite{Epple:2004ra}, such corrections were computed in several orientifolds of toroidal orbifolds with intersecting D6-branes/O6-planes and found to arise from annulus, M{\"{o}}bius strip and Klein bottle diagrams.
The corrections are proportional to the intersection volume and appear at order $g_s^2\, \alpha^{\prime 3}$, which indeed agrees with the order obtained by evaluating \eqref{dsgusgh} for $p=p^\prime=6$. This suggests that the corrections of \cite{Epple:2004ra} could be concrete realizations of our mechanism. Note that this would also make it unlikely that our correction can be removed by a field redefinition.
In view of the incomplete knowledge of perturbative corrections to the D-brane/O-plane effective action, we leave a verification of this claim for future work.

\subsection{Intersecting D7-branes}

As explained above, the coefficient of volume corrections of the form \eqref{dsgusgh} is not fixed by our general arguments, and the corrections could, at least in some cases, disappear by field redefinitions, which is difficult to check directly given the rather fragmentary knowledge of $\alpha^\prime$ corrections to the DBI action. In case the corrections are a coordinate artifact, they have of course no impact on moduli stabilization. On the other hand, scattering amplitudes in orbifold models with intersecting D-branes do lead to corrections of the above form. In the following, we will therefore take the above corrections at face value and analyze how dangerous their presence could potentially be for moduli stabilization. In order to illustrate our general mechanism with an example, we discuss the effect of a correction in type IIB/F-theory compactifications with intersecting D$7$-branes and O$7$-planes, where it could appear already at order $\alpha^{\prime 2}g_s^2$ in the K{\"{a}}hler potential and thus enter as the leading correction to its classical no-scale behavior.

In the absence of D$3$-branes and fluxes, there is no source for the warp factor in 10D Einstein frame such that the Einstein frame metric is unwarped. However, as is well known, D$7$-branes and O$7$-planes generate a non-trivial profile for the dilaton. The string frame metric is therefore warped,
\begin{equation}
\d s_{10}^2 = \e^{\phi(y)/2} \tilde g_{\mu\nu} \d x^\mu \d x^\nu + \e^{\phi(y)/2} \tilde g_{mn} \d y^m \d y^n,
\end{equation}
where the tilded metric coincides with the 10D Einstein frame metric. The presence of warping in the string frame metric can also be understood from the point of view of the M-theory/F-theory duality \cite{Denef:2008wq}. On the M-theory side, the circle along which ones T-dualizes is non-trivially fibered over the base manifold. After T-duality, this direction decompactifies and thus becomes part of the external spacetime on the type IIB side. The external spacetime then depends non-trivially on the internal coordinates and is therefore warped.

The above metric corresponds to \eqref{warpedmetric} for the special case $A=B=(\phi-\phi_0)/4$. Substituting this choice for $A$ and $B$ into \eqref{induced-eh3}, we find that the volume shift due to a single D$7$-brane is
\begin{align}
\Delta \mathcal{V} = \frac{5}{12\cdot 64} (2\pi \alpha^{\prime})^2\, 2\kappa^2 g_s\, \mu_7 \int \d^4 y\, \sqrt{\tilde g^{(4)}}\, (\tilde{\partial_T \phi})^2. \label{induced-eh4}
\end{align}
In the weak coupling limit, the dilaton is constant along the worldvolume except at loci where the brane intersects with another D$7$-brane or O$7$-plane and the derivative of the dilaton jumps. This can easily be verified by putting a probe D$7$-brane into a background with another D$7$-brane or O$7$-plane that is localized along the worldvolume of the first brane. The equations of motion then yield
\begin{equation}
\tilde \nabla_T^2 \phi - (\tilde{\partial_T \phi})^2 = 2 \kappa^2\mu_7\, \e^{\phi} \tilde \delta^{(2)}(y-y_0).
\end{equation}
Using this together with $2\kappa^2 \mu_7 = 1$ in \eqref{induced-eh4}, we find that the volume shift induced on a single probe D$7$-brane in a background generated by a second D$7$-brane is
\begin{equation}
\Delta \mathcal{V} = -\frac{5}{12\cdot 64} (2\pi \alpha^{\prime})^2 g_s^2\, \mathcal{V}_{\textrm{D}7 \cap \textrm{D}7},
\end{equation}
where $g_s = \e^{\phi_0}$ is the string coupling and $\mathcal{V}_{\textrm{D}7 \cap \textrm{D}7} = \int \d^2 y \sqrt{\tilde g_2} \,\e^{\phi-\phi_0}$ is the warped intersection volume. Analogously, the total volume shift due to the curvature corrections of intersecting D$7$-branes and O$7$-planes is proportional to their intersection volume,
\begin{equation}
\Delta \mathcal{V} \propto \alpha^{\prime 2} g_s^2\, \mathcal{V}_{\textrm{D}7 \cap \textrm{O}7}. \label{induced-eh5}
\end{equation}

The numerical proportionality factor in \eqref{induced-eh5} depends on the number of D$7$-branes and O$7$-planes present in the compactification as well as their particular intersection pattern. In order to go beyond a probe estimate of the parametric dependence, one has to compute the correction in a consistent compactification including the full backreaction of the various sources. The most convenient framework to do this should be an F-theory computation, as the backreaction of the 7-branes is then automatically incorporated by choosing a particular Calabi-Yau 4-fold. This also allows to consider brane configurations that are difficult to analyze in the type IIB picture. In particular, $7$-branes in general F-theory solutions do not wrap smooth surfaces but are known to recombine into self-intersecting singular surfaces such as the so-called Whitney branes \cite{Collinucci:2008pf}.

Contrary to the case of intersecting D6-branes/O6-planes, explicit models with one-loop corrections of the form \eqref{induced-eh5} have, to our knowledge, not been discussed in the literature so far. In \cite{Berg:2005ja}, the authors computed loop corrections to the K{\"{a}}hler potential in several examples of toroidal orientifolds with D3-branes/O3-planes and intersecting D7-branes/O7-planes. They found that the volume dependence of the K{\"{a}}hler potential is corrected by terms arising from an exchange of KK modes between the D3-branes/O3-planes and the D7-branes/O7-planes. Furthermore, it was found in \cite{Berg:2005ja} that winding strings on intersections of D7-branes and/or O7-planes can give further corrections to the K{\"{a}}hler potential of $\mathcal{N}=1$ compactifications. These corrections are, however, not related to the one discussed in this section since their volume dependence is different from ours. This makes sense, as we do not expect to capture winding effects within our supergravity approach. However, it was mentioned in \cite{Berg:2007wt} that one-loop corrections from KK mode exchange are in general also expected to arise between D7-branes. Moreover, it was argued there that such corrections should also be present in smooth Calabi-Yau compactifications, and a general form of these corrections was proposed (see also \cite{Cicoli:2007xp}), where the proposed volume scaling indeed matches with the one of our correction. It would be very interesting to confirm this in explicit models.

Let us now comment on the expected effects for moduli stabilization in a possible compactification with corrections of the form \eqref{induced-eh5}. Expanding the K{\"{a}}hler potential \eqref{kaehler} at large volume, one finds a leading order logarithmic behavior plus an infinite sum of higher order terms that are suppressed by inverse powers of the volume. The leading contribution of \eqref{induced-eh5} to such an expansion goes like $\mathcal{V}_{\textrm{D}7 \cap \textrm{O}7}/\mathcal{V}$ and is therefore of degree $-2$ in the (classical) 2-cycle volumes. Naively, it is then of lower order in the volume expansion than the usual tree-level $\alpha^{\prime 3}$ correction \cite{Becker:2002nn}, which goes like $1/\mathcal{V}$ and is therefore of degree $-3$. However, it was realized in \cite{Berg:2007wt, Cicoli:2007xp} that corrections of degree $-2$ to the K{\"{a}}hler potential obey a so-called extended no-scale structure. This means that, even though they are leading in the K{\"{a}}hler potential compared to the $\alpha^{\prime 3}$ correction, a subtle cancellation ensures that they contribute to the scalar potential only at subleading order. The effect of a one-loop correction of the form \eqref{induced-eh5} on moduli stabilization would thus be smaller than expected from a naive counting of the powers of $\alpha^\prime$.
Considering this together with the additional suppression of the correction by $g_s$, we do therefore not expect dramatic effects of the kind that could have arisen \cite{Pedro:2013qga} in the presence of the tree-level corrections proposed in \cite{Grimm:2013gma, Grimm:2013bha}.

\subsection{Other corrections}

The main purpose of this section was to emphasize that corrections to the K{\"{a}}hler potential due to brane intersections are expected to appear at one loop. As we have indicated above, however, these are in general not the only effects at this order in the $g_s$ expansion. Further corrections at the same order could descend, e.g., from the dimensional reduction of bulk corrections to the 10D effective action, i.e., from torus diagrams. By construction, such corrections are not captured by the above analysis, which focussed on corrections due to branes. Furthermore, as discussed in Section \ref{action}, the knowledge of brane actions is incomplete already at next-to-leading order such that additional corrections are expected to appear also in the dimensionally-reduced effective action. Finally, numerous corrections appear at other orders in $\alpha^\prime$ and $g_s$ such as, e.g., the winding corrections described in \cite{Berg:2005ja}.
For phenomenological applications, it would obviously be important to get a more complete picture of all of these corrections and to check to which extent they affect moduli stabilization, analogously to what we discussed above. An exhaustive treatment of all possible brane and bulk corrections is beyond current technology such that we leave this for future work.

\section{K{\"{a}}hler corrections from induced D3-brane charge}
\label{d3}

In the previous sections, we have argued that corrections to the K{\"{a}}hler potential that shift the classical volume appear earliest at one-loop order in type II string theory. Nevertheless, corrections that modify the volume dependence of the K{\"{a}}hler potential already at tree-level do exist, however, they do not shift the classical volume itself but instead only redefine the K{\"{a}}hler coordinates in terms of the volume. In type IIB compactifications, this is well-known to happen in the presence of a D$3$-brane and is due to a non-trivial fibration of the $C_4$ axion moduli space over the moduli space of the worldvolume scalar fields that describe the positions of the D$3$-brane in the transverse space. In this section, we review how this correction comes about and point out that it more generally also appears in the presence of D$7$-branes due to curvature corrections to their WZ action.
In the latter case, the correction can be understood from an induced D$3$-brane charge on the worldvolume of the D$7$-branes.

\subsection{K{\"{a}}hler coordinates in the presence of D3-branes}
\label{d3-review}

It is well-known that the 4D effective action describing $\mathcal{N}=1$ type IIB/F-theory compactifications with D$3$-branes exhibits a non-trivial mixing of the worldvolume moduli with the bulk K{\"{a}}hler coordinates in the K{\"{a}}hler potential. It was suggested in \cite{DeWolfe:2002nn} that this mixing is of the schematic form
\begin{equation}
K = - 3 \ln \left(T+\bar T - k(\bar\Phi\Phi)\right) + \ldots \label{schematic-kaehler}
\end{equation}
in an appropriate normalization, where $K$ is the K{\"{a}}hler potential, $T$ are the K{\"{a}}hler coordinates and $k$ is a function of the complex scalar fields $\Phi$, which live on the worldvolumes of the D$3$-branes and determine their positions in the compact space. This form of the K{\"{a}}hler potential was explicitly confirmed in \cite{Grana:2003ek}, where the authors derived the 4D effective action of (unwarped) $\mathcal{N}=1$ Calabi-Yau orientifold compactifications with D$3$-branes and fluxes via a dimensional reduction of the type IIB supergravity action. Warping corrections to this K{\"{a}}hler potential were obtained in \cite{Chen:2009zi}. It was furthermore shown in \cite{Grana:2003ek, Chen:2009zi} that the mixing \eqref{schematic-kaehler} is due to the coupling of a D$3$-brane to the RR potential $C_4$ in the Wess-Zumino part of the brane action. 

This suggests that not only D$3$-branes but, more generally, objects carrying D$3$-brane charge should lead to such a mixing. Indeed, it was found in \cite{Grana:2003ek} that this happens in the presence of non-zero 3-form flux. Similar effects also arise in $\mathcal{N}=1$ Calabi-Yau orientifolds with D$7$-branes, which was shown in \cite{Jockers:2004yj, Jockers:2005zy} for the unwarped limit (see also \cite{Jockers:2005pn} for a detailed review) and confirmed in \cite{Marchesano:2008rg} taking into account warping corrections to the effective action. The worldvolume scalars then mix with the axio-dilaton in the K{\"{a}}hler potential, which can be traced back to the coupling of the D$7$-branes to $C_8$. In the presence of a non-zero $B$ field or worldvolume gauge flux, on the other hand, the Wess-Zumino action of a D$7$-brane also contains couplings to $C_4$ such that the brane can carry an induced D$3$-brane charge. Fluctuations of the $B$ field and Wilson line moduli then again lead to shifts in the K{\"{a}}hler coordinates as in \eqref{schematic-kaehler}.

As this observation is essential for the main point of this section, let us briefly review how this comes about. In order to illustrate our point, we will consider the simple example of D$3$-branes in a Calabi-Yau orientifold background (as the discussion is considerably simpler than the D$7$-brane cases which we will turn to in the next section). We will closely follow the
analysis of \cite{Grana:2003ek} in our discussion.

In order to determine how the presence of D$3$-branes affects the 4D effective action that governs the low-energy dynamics of the compactification, we have to dimensionally reduce the brane action \eqref{dpaction} for $p=3$. The standard way to do this is to allow the 10D bulk fields that couple to the branes to fluctuate around their background configuration and then restrict to the zero modes by expanding the fluctuations in terms of harmonic forms on the compact space. The resulting expression is then pulled back to the worldvolume including the brane deformations $\Phi^m$ and substituted into \eqref{dpaction}.

For simplicity, let us set to zero the $B$ field and the worldvolume gauge field strength in the following. The D$3$-brane action then reduces to
\begin{equation}
S_{\textrm{D}3} = - \mu_3 \int_\mathcal{W} \d^{4} \xi \,\e^{-\phi} \,\mathrm{Tr} \sqrt{-\det (P[g_{\mu\nu}]) } + \mu_3 \int_\mathcal{W} \mathrm{Tr}\, P[C_4]. \label{d3action}
\end{equation}
Furthermore, we will consider a model with a single K{\"{a}}hler modulus and neglect possible contributions to the K{\"{a}}hler potential due to bulk fluxes. The full expressions for the general case are given in \cite{Grana:2003ek} but not required for the present discussion.

Let us at first discuss the dimensional reduction of the WZ action. We start by splitting the RR potential $C_4$ into a background and a fluctuation, $C_4 = \langle C_4\rangle + \delta C_4$. Following \cite{Grana:2003ek}, we then expand
\begin{equation}
\delta C_4(x) = D_{2} (x) \w \omega_2 + V_1^{a}(x) \w \alpha_{3a} + U_{1a}(x) \w \beta_3^{a} + \rho(x) \w \omega_4, \label{c4-expansion}
\end{equation}
where $\omega_2$ is a harmonic 2-form, $\alpha_{3a}$ and $\beta_3^{a}$ are harmonic 3-forms and $\omega_4$ is a harmonic 4-form on the Calabi-Yau orientifold, respectively. The coefficients $V_1^{a}$ and $U_{1a}$ are then 1-form fields, $D_{2}$ is a 2-form field and $\rho$ is a scalar field in the 4D effective field theory. Due to the self-duality of $F_5$, two of the fields, e.g., $V_1^{a}$ and $D_{2}$, are actually redundant and can be eliminated in terms of the other two fields via Poincar{\'{e}} duality.

The relevant term for the mixing \eqref{schematic-kaehler} in the K{\"{a}}hler potential is the first term on the right-hand side of \eqref{c4-expansion}. This term can now be pulled back to the worldvolume, where we allow small fluctuations of the branes away from their background position. One can show that this yields \cite{Grana:2003ek}
\begin{equation}
P[C_4] = \langle C_4 \rangle + \frac{1}{4} (2\pi\alpha^\prime)^2 (\Phi^m D_\mu \bar\Phi^{\bar n} - \bar\Phi^{\bar n} D_\mu \Phi^m) \omega_{m \bar n} \d x^\mu \w \d D_{2} + \ldots \label{b}
\end{equation}
up to terms involving higher powers of the worldvolume scalars $\Phi^m$, where $D_\mu$ is the gauge covariant derivative. Since the worldvolume scalars parametrize deformations normal to the wordvolume, they are properly defined as sections of the normal bundle \cite{Bershadsky:1995qy}. $D_\mu$ therefore also contains a connection to the normal bundle.

We now substitute \eqref{b} back into \eqref{d3action} and eliminate the 2-form field $D_{2}$ in terms of the dual scalar $\rho$. This leads to a mixed kinetic term of the form
\begin{equation}
\frac{\mu_3}{4} (2\pi\alpha^\prime)^2 \left(\mathrm{Tr} (\Phi^m D_\mu \bar\Phi^{\bar n} - \bar\Phi^{\bar n} D_\mu \Phi^m) \omega_{m \bar n} \right)\left(\partial^\mu \rho\right) \label{mixkinetic}
\end{equation}
in the 4D effective action. Such a mixing was also observed in \cite{Frey:2002hf}.

Let us now consider the DBI part of the action \eqref{d3action}. Similar to what we have done above for $C_4$, we now expand the internal metric into fluctuations,
\begin{equation}
g_{m \bar n} = -i \mathcal{V}(x)^{1/3} \omega_{m \bar n},
\end{equation}
where $\mathcal{V}(x)$ is the volume modulus and we have neglected the complex structure deformations as they are not relevant for the present discussion. The pullback of the metric to the brane worldvolume is then again given by a normal coordinate expansion,
\begin{equation}
P[g_{\mu\nu}] = \langle g_{\mu\nu} \rangle -2 i (2\pi\alpha^\prime)^2 \,\mathcal{V}^{1/3} \omega_{m \bar n} (D_\mu \Phi^m)(D_\nu \bar \Phi^{\bar n}) + \ldots
\end{equation}
One readily observes, that, in contrast to \eqref{b}, the expansion does not yield any mixed kinetic terms between the worldvolume scalars $\Phi^m$ and the fluctuations of the pulled-back bulk field, i.e., the volume modulus $\mathcal{V}$. Such a mixing cannot appear since, by definition of the normal coordinate expansion, the derivative $D_\mu$ is covariant with respect to the metric $g_{m \bar n} = -i \mathcal{V}^{1/3} \omega_{m \bar n}$ and does therefore not act on it.

Hence, the dimensional reduction of \eqref{d3action} yields a mixed kinetic term between the worldvolume scalars $\Phi^m$ and the imaginary part $\rho$ of the K{\"{a}}hler coordinate $T$ but no mixed kinetic term involving its real part $\mathcal{V}$. In $\mathcal{N}=1$ supergravity, the structure of the kinetic terms of the theory is fully determined by the K{\"{a}}hler metric $g_{i \bar \jmath}$,
\begin{equation}
\mathcal{L}_\textrm{kin} =  g_{i \bar \jmath} (\phi^i, \bar \phi^{\bar \jmath}) (D_\mu \phi^i)(D^\mu \bar \phi^{\bar \jmath}), \qquad g_{i \bar \jmath} = \frac{\partial^2 K(\phi^i, \bar \phi^{\bar \jmath})}{\partial \phi^i \partial \bar \phi^{\bar \jmath}}, \label{lkin}
\end{equation}
where $\phi^i = \left\{ T, \Phi^m, \ldots \right\}$. In the absence of D$3$-branes, the K{\"{a}}hler potential takes the usual no-scale form \cite{Grana:2003ek}
\begin{equation}
K = -3 \ln \left( T + \bar T \right) + \ldots, \qquad T = \frac{3i}{2} \rho + \frac{9}{2} \mathcal{V}^{2/3},
\end{equation}
where the dots indicate contributions from other sectors such as the axio-dilaton and the complex structure moduli. It is then easy to check that the mixed term \eqref{mixkinetic} is reproduced by \eqref{lkin} if the K{\"{a}}hler potential is modified\footnote{ Note that factors of the 4D Planck mass were set to 1 in the expressions of \cite{Grana:2003ek}.} such that
\begin{equation}
K = -3 \ln \left( T + \bar T - 3i\mu_3(2\pi\alpha^\prime)^2 \omega_{m \bar n} \mathrm{Tr} (\Phi^m \bar\Phi^{\bar n}) \right) + \ldots \label{shift1}
\end{equation}
and the K{\"{a}}hler coordinates are shifted as
\begin{equation}
T = \frac{3i}{2} \rho + \frac{9}{2} \mathcal{V}^{2/3} + \frac{3i}{2}\mu_3 (2\pi\alpha^\prime)^2 \omega_{m \bar n} \mathrm{Tr} (\Phi^m \bar\Phi^{\bar n}). \label{shift2}
\end{equation}
It should be stressed here that both shifts \eqref{shift1} and \eqref{shift2} are crucial in order to recover the structure of the kinetic terms that we derived from the dimensional reduction. If only the K{\"{a}}hler potential were shifted without redefining the K{\"{a}}hler coordinates, \eqref{lkin} would yield a mixed kinetic term between $\mathcal{V}$ and $\Phi^m$, which is, as we explained above, not present in the effective action.

For large fluctuations $\Phi^m$, the above expressions receive further corrections due to higher power terms such that the shift will in general be proportional to some function of $\Phi^m$, whose explicit form depends on the considered background. For the special case of an unwarped Calabi-Yau orientifold, some higher order corrections were computed in \cite{Grana:2003ek}. Schematically denoting the shift by $k(\bar \Phi\Phi)$, we find
\begin{equation}
K = - 3 \ln \left(T+\bar T - k(\bar\Phi \Phi)\right) + \ldots, \qquad T = \frac{3i}{2} \rho + \frac{9}{2} \mathcal{V}^{2/3} + \frac{1}{2} k(\bar\Phi \Phi).
\end{equation}
One furthermore checks that the K{\"{a}}hler potential can be rewritten as
\begin{equation}
K = -2 \ln \mathcal{V} + \ldots
\end{equation}
Hence, the physical volume stays invariant under inserting and shifting the positions of the D$3$-branes.

\subsection{Induced D3-brane charge on D7-branes}

\label{d7}

As we have reviewed in the previous subsection, objects that carry a D$3$-brane charge lead to a shift in the K{\"{a}}hler coordinates of the 4D effective action. In this section, we argue that this reasoning also applies to D$7$-branes due to the curvature corrections to their WZ action, which induce a D3-brane charge on their worldvolume.

As discussed in Section \ref{action}, the curvature corrections depend on the curvature 2-forms of the tangent and normal bundles of the worldvolume and can be written in terms of characteristic classes. For simplicity, let us assume a vanishing $B$ field and worldvolume gauge field strength in the following.
Evaluating \eqref{aa} for a D7-brane wrapping a 4-cycle $S$, we then find a coupling to $C_4$ and thus an induced D$3$-brane charge
\begin{equation}
Q_3^{\textrm{D}7} = \frac{\mu_3}{48} \int_S \left( p_1(NS) - p_1(TS) \right),
\end{equation}
where we have used $\mu_3 = \mu_7 (2\pi)^4\alpha^{\prime 2}$. Let us assume for now that $S$ is a smooth complex hypersurface in a complex 3-dimensional manifold $X_3$. Using standard formulae of algebraic geometry, one can then show that, if $X_3$ is Calabi-Yau (i.e., $c_1(X_3)=0$), the above expression reduces to
\begin{equation}
Q_3^{\textrm{D}7} = \frac{\mu_3}{24} \int_S c_2(TS) = \frac{\mu_3}{24} \, \chi(S), \label{induced1}
\end{equation}
where $\chi(S)$ is the Euler characteristic of $S$. An analogous argument leads to the induced D3-brane charge of an O$7^-$-plane wrapped on a 4-cycle $S^\prime$,
\begin{equation}
Q_3^{\textrm{O}7^-} = \frac{\mu_3}{12} \int_{S^\prime} \left( p_1(NS^\prime) - p_1(TS^\prime) \right) = \frac{\mu_3}{6} \, \chi(S^\prime). \label{induced2}
\end{equation}

\eqref{induced1} and \eqref{induced2} are correct for type IIB D7-branes and O7-planes wrapping smooth hypersurfaces in a Calabi-Yau 3-fold. Naively, one might expect that the same is true in F-theory compactifications if one considers Sen's weak coupling limit \cite{Sen:1996vd, Sen:1997gv}, where the F-theory background has a description in terms of a perturbative type IIB Calabi-Yau orientifold with $c_1(X_3)=0$ (where $X_3$ is the double cover of the base $B_3$ of the elliptic fibration). For generic brane configurations, however, the above formula does not hold even in the perturbative orientifold limit. The reason is that generic D$7$-branes do not wrap smooth surfaces but self-intersect at the fixed point locus of the orientifold in the shape of a Whitney umbrella \cite{Collinucci:2008pf} (see also \cite{Braun:2008ua}). This has the effect that there are generically singularities on the base $B_3$ (and on the hypersurface $S$) such that the correct evaluation of the induced D$3$-brane charge can be subtle \cite{Collinucci:2008pf, Sethi:1996es}. The induced D$3$-brane charge is then given by a modified Euler characteristic $\chi_o(S)$, which appropriately generalizes the notion of an Euler characteristic to singular algebraic varieties \cite{Collinucci:2008pf},
\begin{equation}
Q_3^{\textrm{D}7} = \frac{\mu_3}{24} \, \chi_o(S), \qquad Q_3^{\textrm{O}7^-} = \frac{\mu_3}{6} \, \chi(S^\prime). \label{d3charge-weak}
\end{equation}

In non-perturbative F-theory backgrounds at strong coupling, the backreaction of the $7$-branes deforms the internal geometry of $X_3$ away from a Calabi-Yau such that the description in terms of a type IIB orientifold is not valid anymore. The total curvature-induced D$3$-brane charge of D$7$-branes, O$7$-planes and (if present) O$3$-planes is then given by the Euler characteristic of the Calabi-Yau 4-fold $X_4$. Alternatively, it can be written in terms of characteristic classes on the base $B_3$ \cite{Sethi:1996es, Klemm:1996ts}.

\subsection{K{\"{a}}hler potential}

Let us now discuss how the above is related to corrections to the K{\"{a}}hler potential. As we have seen in the previous subsections, curvature corrections to the D$7$-brane action contain terms of the form
\begin{equation}
\frac{\mu_3}{48} \int_\mathcal{W} P[C_4 \w X_4],
\end{equation}
where $X_4$ is a 4-form for a correction localized on the D$7$ worldvolume. Note that these terms have exactly the same form as the $C_4$ coupling of a D$3$-brane except for a different numerical prefactor and a different worldvolume dimension. Hence, one again expects to find a correction to the K{\"{a}}hler potential of the form
\begin{equation}
K = - 3 \ln \left(T+\bar T - k(\bar\zeta \zeta)\right), \qquad T = \frac{3i}{2} \rho + \frac{9}{2} \mathcal{V}^{2/3} + \frac{1}{2} k(\bar\zeta \zeta), \label{d7shift1}
\end{equation}
where $\zeta$ parametrizes the deformations of the D$7$-brane. Repeating the steps discussed in Section \ref{d3-review} and using general formulae for the pullback of $p$-form fields \cite{Jockers:2004yj}, we find
\begin{equation}
k(\bar\zeta\zeta) \propto i \mu_3 \alpha^{\prime 2} \chi(S)\, \omega_{ m \bar n} \zeta^m \bar\zeta^{\bar n} \label{d7shift}
\end{equation}
up to terms involving higher powers of the fluctuations $\zeta^m$.

For simplicity, we have assumed here that the deformations of the worldvolume are parametrized by fields transforming as sections of the normal bundle $NS$. It was argued in \cite{Beasley:2008dc} that, for general curved $S$, the worldvolume ``scalars'' $\zeta$ rather transform as sections of the canonical bundle $K_S$ and are therefore $(2,0)$ forms on the worldvolume (see also \cite{Bershadsky:1995qy}). In a type II orientifold, they are odd under the orientifold involution and are therefore elements of the cohomology class $H^{2,0}_-(S)$ \cite{Jockers:2004yj}. For a brane wrapping a smooth surface on a Calabi-Yau threefold, both descriptions can actually be shown to be equivalent \cite{Jockers:2004yj}, but this is not true anymore for generic $7$-branes in F-theory, as discussed in detail in \cite{Beasley:2008dc}. Even in the weakly coupled regime, where $X_3$ is a Calabi-Yau, the presence of singularities then makes the above parametrization too naive since the normal bundle is then not properly defined at the intersection. It was shown in \cite{Collinucci:2008pf} that the proper way to count the brane deformations is then to resolve the singular surface $S$ to $\Sigma$ and compute the cohomology class $H^{2,0}_-(\Sigma)$. The above expression thus has to be modified accordingly in these more general setups.

A well-known feature of brane intersections is that they give rise to localized matter fields, i.e., additional worldvolume scalars localized at the intersection locus. For a D$7$-brane that fills the 4D external spacetime and self-intersects on an algebraic curve in the compact space, this locus is 6-dimensional.
In F-theory, localized matter can arise on intersections of $7$-branes with more general gauge groups than those of perturbative D$7$-branes \cite{Beasley:2008dc, Donagi:2008ca}. It would be interesting to study the above corrections for these more general brane configurations and analyze their implications for phenomenology.

\subsection{Other corrections}

Like the corrections of Section \ref{eh}, the corrections found here are not expected to be the only ones of their form at the same order in $\alpha^\prime$ and $g_s$. Rather, they are an example of how the volume dependence of the K{\"{a}}hler potential can be modified by D-branes already at tree-level in $g_s$. More generally, however, any term involving a coupling to $C_4$ induces a D3-brane charge and is therefore expected to correct the volume dependence of the K{\"{a}}hler potential by the mechanism discussed in Section \ref{d3-review}. Such corrections can either come from higher-order terms in the 10D bulk effective action or from brane corrections other than the $C_4 R^2$ couplings discussed above. An example are some of the T-duality covariantized couplings derived in \cite{Becker:2010ij}.

\section{Conclusions}
\label{conclusions}

Motivated by the recurrent role of $\alpha'$ effects in string phenomenology, we have initiated a study of higher-derivative corrections to the 4D K{\"{a}}hler potential of $\mathcal{N}=1$ type II string compactifications arising from curvature terms in the DBI and WZ actions of D-branes and O-planes. We argued that these corrections can be present in generic string vacua with localized sources and thus should be taken into account in conjunction with the (familiar)  bulk curvature corrections to supergravity. Of particular phenomenological interest is the volume modulus, and so we have identified two mechanisms in which these brane curvature terms can correct its 4D K{\"{a}}hler potential. The first mechanism comes from brane curvature corrections inducing an effective Einstein-Hilbert term on warped brane worldvolumes, and thus the classical volume of the compactification
manifold is shifted at one-loop order. This mechanism applies to setups with or without brane intersections. In the case of intersecting D7-branes/O7-planes, we showed that possible corrections are proportional to the intersection volume and can appear already at ${\cal O} (\alpha'^2 g_s^2)$ in the K{\"{a}}hler potential. Assuming that such corrections cannot be removed by a field redefinition, they could then potentially be dangerous for moduli stabilization. We argued, however, that their presence would only give subleading effects to the scalar potential in comparison to the well-known ${\cal O} (\alpha'^3)$ corrections due to the presence of an extended no-scale structure.
The second mechanism we discussed in this paper is due to an induced D3-brane charge on the worldvolume of D7-branes. Contrary to the first type of corrections, this effect appears at open string tree-level. The corresponding correction shifts the definition of the K{\"{a}}hler coordinates in terms of the classical volume but leaves the volume itself uncorrected.
Along the way, we also showed that the ${\cal O} (\alpha^{\prime 2})$ corrections proposed in  \cite{Grimm:2013gma} are really an artifact of an inconvenient basis of M-theory fields upon duality transformation and can be removed by a field redefinition at the 10D/11D level. We thus convincingly conclude that the leading corrections to the classical volume from D-branes and O-planes can appear earliest at the one-loop order and are therefore not expected to compromise the usual moduli stabilization scenarios.

There are several promising avenues for our results to be generalized and applied. While we found the parametric dependence of the one-loop corrections for the volume modulus in Section \ref{eh}, it would certainly be of interest to compute the coefficients of such corrections in explicit models and explore in more detail their phenomenological consequences. For example, one should revisit their implications for moduli stabilization scenarios \cite{Pedro:2013qga} in light of the different $g_s$ dependence we found. A direct worldsheet analysis of the corrections we found here would provide further support of our results \cite{future}. Other than moduli stabilization, $\alpha'$ effects are important for addressing issues in particle physics and cosmology that are highly UV sensitive, e.g., gravity-mediated supersymmetry breaking and inflation. While inflation is generically sensitive to leading Planck suppressed corrections, the UV sensitivity is even stronger for large field inflation. The recent BICEP2 result, if confirmed to be primordial in origin, would put more pressure in understanding string and quantum gravity corrections to supergravity as it suggests an inflationary energy scale $V^{1/4} \sim 10^{16}$ GeV and a super-Planckian inflaton field excursion. We hope to return to some of these phenomenological applications in the future.

On a related note, it is somewhat surprising how little is known about higher-order corrections to the D-brane and O-plane effective actions. Our work highlighted the significance of these corrections. It would be important, from both a formal and a phenomenological point of view, to obtain the fully T-duality covariantized curvature corrections at order $\alpha^{\prime 2}$ for both D-branes and O-planes. Some steps in this direction have recently been undertaken in \cite{Robbins:2014ara}. More generally, it would be desirable to get a more complete picture of the leading $\alpha^\prime$ and $g_s$ corrections to the effective action of type IIB/F-theory compactifications in order to further improve on our understanding of moduli stabilization at large volume and small string coupling. An important long-term goal is of course to explore the landscape beyond the perturbative regime, possibly by further developing techniques in the spirit of \cite{Sen:2013oza}. We hope to report on our progress in some future work.

\section*{Acknowledgements}

The authors would like to thank William Cottrell for numerous discussions on the subject. We also benefitted from discussions with Costas Bachas, Andr{\'{e}}s Collinucci, I{\~{n}}aki Garc{\'{i}}a Etxebarria, Thomas Grimm, Jan Hajer, Fernando Marchesano, Raffaele Savelli, Pablo Soler Gomis and Matthias Wei\ss enbacher. This work is supported in part by the DOE grant DE-FG-02-95ER40896 and the HKRGC grants 604213 and HKUST4/CRF/13G. DJ also thanks the University of Wisconsin, Madison for hospitality during a visit where part of this work was completed.

\bibliographystyle{utphys}
\bibliography{groups}

\providecommand{\href}[2]{#2}\begingroup\raggedright\begin{thebibliography}{100}

\bibitem{Kachru:2003aw}
S.~Kachru, R.~Kallosh, A.~D. Linde and S.~P. Trivedi,  {\em {De Sitter vacua in
  string theory}}, Phys.Rev. {\bf D68} (2003) 046005
[\href{http://www.arXiv.org/abs/hep-th/0301240}{{\tt hep-th/0301240}}].

\bibitem{Balasubramanian:2005zx}
V.~Balasubramanian, P.~Berglund, J.~P. Conlon and F.~Quevedo,  {\em
  {Systematics of moduli stabilisation in Calabi-Yau flux compactifications}},
  JHEP {\bf 0503} (2005) 007
[\href{http://www.arXiv.org/abs/hep-th/0502058}{{\tt hep-th/0502058}}].

\bibitem{Conlon:2005ki}
J.~P. Conlon, F.~Quevedo and K.~Suruliz,  {\em {Large-volume flux
  compactifications: Moduli spectrum and D3/D7 soft supersymmetry breaking}},
  JHEP {\bf 0508} (2005) 007
[\href{http://www.arXiv.org/abs/hep-th/0505076}{{\tt hep-th/0505076}}].

\bibitem{Balasubramanian:2004uy}
V.~Balasubramanian and P.~Berglund,  {\em {Stringy corrections to K{\"{a}}hler
  potentials, SUSY breaking, and the cosmological constant problem}}, JHEP {\bf
  0411} (2004) 085
[\href{http://www.arXiv.org/abs/hep-th/0408054}{{\tt hep-th/0408054}}].

\bibitem{Westphal:2006tn}
A.~Westphal,  {\em {de Sitter string vacua from K{\"{a}}hler uplifting}}, JHEP
  {\bf 0703} (2007) 102
[\href{http://www.arXiv.org/abs/hep-th/0611332}{{\tt hep-th/0611332}}].

\bibitem{Rummel:2011cd}
M.~Rummel and A.~Westphal,  {\em {A sufficient condition for de Sitter vacua in
  type IIB string theory}}, JHEP {\bf 1201} (2012) 020
[\href{http://www.arXiv.org/abs/1107.2115}{{\tt 1107.2115}}].

\bibitem{Louis:2012nb}
J.~Louis, M.~Rummel, R.~Valandro and A.~Westphal,  {\em {Building an explicit
  de Sitter}}, JHEP {\bf 1210} (2012) 163
[\href{http://www.arXiv.org/abs/1208.3208}{{\tt 1208.3208}}].

\bibitem{Berg:2005yu}
M.~Berg, M.~Haack and B.~K{\"{o}}rs,  {\em {On volume stabilization by quantum
  corrections}}, Phys.Rev.Lett. {\bf 96} (2006) 021601
[\href{http://www.arXiv.org/abs/hep-th/0508171}{{\tt hep-th/0508171}}].

\bibitem{Green:2011cn}
S.~R. Green, E.~J. Martinec, C.~Quigley and S.~Sethi,  {\em {Constraints on
  String Cosmology}}, Class.Quant.Grav. {\bf 29} (2012) 075006
[\href{http://www.arXiv.org/abs/1110.0545}{{\tt 1110.0545}}].

\bibitem{Gautason:2012tb}
F.~F. Gautason, D.~Junghans and M.~Zagermann,  {\em {On Cosmological Constants
  from alpha'-Corrections}}, JHEP {\bf 1206} (2012) 029
[\href{http://www.arXiv.org/abs/1204.0807}{{\tt 1204.0807}}].

\bibitem{Dasgupta:2014pma}
K.~Dasgupta, R.~Gwyn, E.~McDonough, M.~Mia and R.~Tatar,  {\em {de Sitter Vacua
  in Type IIB String Theory: Classical Solutions and Quantum Corrections}},
\href{http://www.arXiv.org/abs/1402.5112}{{\tt 1402.5112}}.

\bibitem{Derendinger:2004jn}
J.-P. Derendinger, C.~Kounnas, P.~M. Petropoulos and F.~Zwirner,  {\em
  {Superpotentials in IIA compactifications with general fluxes}}, Nucl.Phys.
  {\bf B715} (2005) 211--233
[\href{http://www.arXiv.org/abs/hep-th/0411276}{{\tt hep-th/0411276}}].

\bibitem{Villadoro:2005cu}
G.~Villadoro and F.~Zwirner,  {\em {N=1 effective potential from dual type-IIA
  D6/O6 orientifolds with general fluxes}}, JHEP {\bf 0506} (2005) 047
[\href{http://www.arXiv.org/abs/hep-th/0503169}{{\tt hep-th/0503169}}].

\bibitem{Camara:2005dc}
P.~G. C{\'{a}}mara, A.~Font and L.~E. Ib{\'{a}}{\~{n}}ez,  {\em {Fluxes, moduli
  fixing and MSSM-like vacua in a simple IIA orientifold}}, JHEP {\bf 0509}
  (2005) 013
[\href{http://www.arXiv.org/abs/hep-th/0506066}{{\tt hep-th/0506066}}].

\bibitem{DeWolfe:2005uu}
O.~DeWolfe, A.~Giryavets, S.~Kachru and W.~Taylor,  {\em {Type IIA moduli
  stabilization}}, JHEP {\bf 0507} (2005) 066
[\href{http://www.arXiv.org/abs/hep-th/0505160}{{\tt hep-th/0505160}}].

\bibitem{Silverstein:2007ac}
E.~Silverstein,  {\em {Simple de Sitter Solutions}}, Phys.Rev. {\bf D77} (2008)
  106006
[\href{http://www.arXiv.org/abs/0712.1196}{{\tt 0712.1196}}].

\bibitem{Caviezel:2008ik}
C.~Caviezel, P.~Koerber, S.~K{\"{o}}rs, D.~L{\"{u}}st, D.~Tsimpis and
  M.~Zagermann,  {\em {The effective theory of type IIA AdS(4)
  compactifications on nilmanifolds and cosets}}, Class.Quant.Grav. {\bf 26}
  (2009) 025014
[\href{http://www.arXiv.org/abs/0806.3458}{{\tt 0806.3458}}].

\bibitem{Caviezel:2008tf}
C.~Caviezel, P.~Koerber, S.~K{\"{o}}rs, D.~L{\"{u}}st, T.~Wrase and
  M.~Zagermann,  {\em {On the Cosmology of Type IIA Compactifications on
  SU(3)-structure Manifolds}}, JHEP {\bf 0904} (2009) 010
[\href{http://www.arXiv.org/abs/0812.3551}{{\tt 0812.3551}}].

\bibitem{Flauger:2008ad}
R.~Flauger, S.~Paban, D.~Robbins and T.~Wrase,  {\em {Searching for slow-roll
  moduli inflation in massive type IIA supergravity with metric fluxes}},
  Phys.Rev. {\bf D79} (2009) 086011
[\href{http://www.arXiv.org/abs/0812.3886}{{\tt 0812.3886}}].

\bibitem{Haque:2008jz}
S.~S. Haque, G.~Shiu, B.~Underwood and T.~Van~Riet,  {\em {Minimal simple de
  Sitter solutions}}, Phys.Rev. {\bf D79} (2009) 086005
[\href{http://www.arXiv.org/abs/0810.5328}{{\tt 0810.5328}}].

\bibitem{Danielsson:2009ff}
U.~H. Danielsson, S.~S. Haque, G.~Shiu and T.~Van~Riet,  {\em {Towards
  Classical de Sitter Solutions in String Theory}}, JHEP {\bf 0909} (2009) 114
[\href{http://www.arXiv.org/abs/0907.2041}{{\tt 0907.2041}}].

\bibitem{Danielsson:2011au}
U.~H. Danielsson, S.~S. Haque, P.~Koerber, G.~Shiu, T.~Van~Riet and T.~Wrase,
  {\em {De Sitter hunting in a classical landscape}}, Fortsch.Phys. {\bf 59}
  (2011) 897--933
[\href{http://www.arXiv.org/abs/1103.4858}{{\tt 1103.4858}}].

\bibitem{Becker:2002nn}
K.~Becker, M.~Becker, M.~Haack and J.~Louis,  {\em {Supersymmetry breaking and
  alpha-prime corrections to flux induced potentials}}, JHEP {\bf 0206} (2002)
  060
[\href{http://www.arXiv.org/abs/hep-th/0204254}{{\tt hep-th/0204254}}].

\bibitem{Berg:2004ek}
M.~Berg, M.~Haack and B.~K{\"{o}}rs,  {\em {Loop corrections to volume moduli
  and inflation in string theory}}, Phys.Rev. {\bf D71} (2005) 026005
[\href{http://www.arXiv.org/abs/hep-th/0404087}{{\tt hep-th/0404087}}].

\bibitem{Berg:2005ja}
M.~Berg, M.~Haack and B.~K{\"{o}}rs,  {\em {String loop corrections to Kahler
  potentials in orientifolds}}, JHEP {\bf 0511} (2005) 030
[\href{http://www.arXiv.org/abs/hep-th/0508043}{{\tt hep-th/0508043}}].

\bibitem{Berg:2007wt}
M.~Berg, M.~Haack and E.~Pajer,  {\em {Jumping Through Loops: On Soft Terms
  from Large Volume Compactifications}}, JHEP {\bf 0709} (2007) 031
[\href{http://www.arXiv.org/abs/0704.0737}{{\tt 0704.0737}}].

\bibitem{Berg:2011ij}
M.~Berg, M.~Haack and J.~U. Kang,  {\em {One-Loop Kahler Metric of D-Branes at
  Angles}}, JHEP {\bf 1211} (2012) 091
[\href{http://www.arXiv.org/abs/1112.5156}{{\tt 1112.5156}}].

\bibitem{Cicoli:2007xp}
M.~Cicoli, J.~P. Conlon and F.~Quevedo,  {\em {Systematics of String Loop
  Corrections in Type IIB Calabi-Yau Flux Compactifications}}, JHEP {\bf 0801}
  (2008) 052
[\href{http://www.arXiv.org/abs/0708.1873}{{\tt 0708.1873}}].

\bibitem{Anguelova:2010ed}
L.~Anguelova, C.~Quigley and S.~Sethi,  {\em {The Leading Quantum Corrections
  to Stringy Kahler Potentials}}, JHEP {\bf 1010} (2010) 065
[\href{http://www.arXiv.org/abs/1007.4793}{{\tt 1007.4793}}].

\bibitem{Vafa:1996xn}
C.~Vafa,  {\em {Evidence for F theory}}, Nucl.Phys. {\bf B469} (1996) 403--418
[\href{http://www.arXiv.org/abs/hep-th/9602022}{{\tt hep-th/9602022}}].

\bibitem{Denef:2008wq}
F.~Denef,  {\em {Les Houches Lectures on Constructing String Vacua}},
\href{http://www.arXiv.org/abs/0803.1194}{{\tt 0803.1194}}.

\bibitem{Grimm:2010ks}
T.~W. Grimm,  {\em {The N=1 effective action of F-theory compactifications}},
  Nucl.Phys. {\bf B845} (2011) 48--92
[\href{http://www.arXiv.org/abs/1008.4133}{{\tt 1008.4133}}].

\bibitem{Weigand:2010wm}
T.~Weigand,  {\em {Lectures on F-theory compactifications and model building}},
  Class.Quant.Grav. {\bf 27} (2010) 214004
[\href{http://www.arXiv.org/abs/1009.3497}{{\tt 1009.3497}}].

\bibitem{Sen:1996vd}
A.~Sen,  {\em {F theory and orientifolds}}, Nucl.Phys. {\bf B475} (1996)
  562--578
[\href{http://www.arXiv.org/abs/hep-th/9605150}{{\tt hep-th/9605150}}].

\bibitem{Sen:1997gv}
A.~Sen,  {\em {Orientifold limit of F theory vacua}}, Phys.Rev. {\bf D55}
  (1997) 7345--7349
[\href{http://www.arXiv.org/abs/hep-th/9702165}{{\tt hep-th/9702165}}].

\bibitem{GarciaEtxebarria:2012zm}
I.~Garc{\'{i}}a-Etxebarria, H.~Hayashi, R.~Savelli and G.~Shiu,  {\em {On
  quantum corrected K{\"{a}}hler potentials in F-theory}}, JHEP {\bf 1303}
  (2013) 005
[\href{http://www.arXiv.org/abs/1212.4831}{{\tt 1212.4831}}].

\bibitem{Grimm:2013gma}
T.~W. Grimm, R.~Savelli and M.~Wei{\ss}enbacher,  {\em {On $\alpha'$
  corrections in N=1 F-theory compactifications}},
\href{http://www.arXiv.org/abs/1303.3317}{{\tt 1303.3317}}.

\bibitem{Grimm:2013bha}
T.~W. Grimm, J.~Keitel, R.~Savelli and M.~Wei{\ss}enbacher,  {\em {From
  M-theory higher curvature terms to $\alpha'$ corrections in F-theory}},
\href{http://www.arXiv.org/abs/1312.1376}{{\tt 1312.1376}}.

\bibitem{Grimm:2012rg}
T.~W. Grimm, D.~Klevers and M.~Poretschkin,  {\em {Fluxes and Warping for Gauge
  Couplings in F-theory}}, JHEP {\bf 1301} (2013) 023
[\href{http://www.arXiv.org/abs/1202.0285}{{\tt 1202.0285}}].

\bibitem{Vafa:1995fj}
C.~Vafa and E.~Witten,  {\em {A One loop test of string duality}}, Nucl.Phys.
  {\bf B447} (1995) 261--270
[\href{http://www.arXiv.org/abs/hep-th/9505053}{{\tt hep-th/9505053}}].

\bibitem{Duff:1995wd}
M.~J. Duff, J.~T. Liu and R.~Minasian,  {\em {Eleven-dimensional origin of
  string-string duality: A One loop test}}, Nucl.Phys. {\bf B452} (1995)
  261--282
[\href{http://www.arXiv.org/abs/hep-th/9506126}{{\tt hep-th/9506126}}].

\bibitem{Green:1997di}
M.~B. Green and P.~Vanhove,  {\em {D instantons, strings and M theory}},
  Phys.Lett. {\bf B408} (1997) 122--134
[\href{http://www.arXiv.org/abs/hep-th/9704145}{{\tt hep-th/9704145}}].

\bibitem{Green:1997as}
M.~B. Green, M.~Gutperle and P.~Vanhove,  {\em {One loop in
  eleven-dimensions}}, Phys.Lett. {\bf B409} (1997) 177--184
[\href{http://www.arXiv.org/abs/hep-th/9706175}{{\tt hep-th/9706175}}].

\bibitem{Kiritsis:1997em}
E.~Kiritsis and B.~Pioline,  {\em {On R**4 threshold corrections in IIb string
  theory and (p, q) string instantons}}, Nucl.Phys. {\bf B508} (1997) 509--534
[\href{http://www.arXiv.org/abs/hep-th/9707018}{{\tt hep-th/9707018}}].

\bibitem{Russo:1997mk}
J.~G. Russo and A.~A. Tseytlin,  {\em {One loop four graviton amplitude in
  eleven-dimensional supergravity}}, Nucl.Phys. {\bf B508} (1997) 245--259
[\href{http://www.arXiv.org/abs/hep-th/9707134}{{\tt hep-th/9707134}}].

\bibitem{Antoniadis:1997eg}
I.~Antoniadis, S.~Ferrara, R.~Minasian and K.~S. Narain,  {\em {R**4 couplings
  in M and type II theories on Calabi-Yau spaces}}, Nucl.Phys. {\bf B507}
  (1997) 571--588
[\href{http://www.arXiv.org/abs/hep-th/9707013}{{\tt hep-th/9707013}}].

\bibitem{Tseytlin:2000sf}
A.~A. Tseytlin,  {\em {R**4 terms in 11 dimensions and conformal anomaly of
  (2,0) theory}}, Nucl.Phys. {\bf B584} (2000) 233--250
[\href{http://www.arXiv.org/abs/hep-th/0005072}{{\tt hep-th/0005072}}].

\bibitem{Liu:2013dna}
J.~T. Liu and R.~Minasian,  {\em {Higher-derivative couplings in string theory:
  dualities and the B-field}},
\href{http://www.arXiv.org/abs/1304.3137}{{\tt 1304.3137}}.

\bibitem{Haack:2001jz}
M.~Haack and J.~Louis,  {\em {M theory compactified on Calabi-Yau fourfolds
  with background flux}}, Phys.Lett. {\bf B507} (2001) 296--304
[\href{http://www.arXiv.org/abs/hep-th/0103068}{{\tt hep-th/0103068}}].

\bibitem{Pedro:2013qga}
F.~G. Pedro, M.~Rummel and A.~Westphal,  {\em {Extended No-Scale Structure and
  $\alpha^{'2}$ Corrections to the Type IIB Action}},
\href{http://www.arXiv.org/abs/1306.1237}{{\tt 1306.1237}}.

\bibitem{Corley:2001hg}
S.~Corley, D.~A. Lowe and S.~Ramgoolam,  {\em {Einstein-Hilbert action on the
  brane for the bulk graviton}}, JHEP {\bf 0107} (2001) 030
[\href{http://www.arXiv.org/abs/hep-th/0106067}{{\tt hep-th/0106067}}].

\bibitem{Bachas:1999um}
C.~P. Bachas, P.~Bain and M.~B. Green,  {\em {Curvature terms in D-brane
  actions and their M theory origin}}, JHEP {\bf 9905} (1999) 011
[\href{http://www.arXiv.org/abs/hep-th/9903210}{{\tt hep-th/9903210}}].

\bibitem{Epple:2004ra}
F.~T.~J. Epple,  {\em {Induced gravity on intersecting branes}}, JHEP {\bf
  0409} (2004) 021
[\href{http://www.arXiv.org/abs/hep-th/0408105}{{\tt hep-th/0408105}}].

\bibitem{Collinucci:2008pf}
A.~Collinucci, F.~Denef and M.~Esole,  {\em {D-brane Deconstructions in IIB
  Orientifolds}}, JHEP {\bf 0902} (2009) 005
[\href{http://www.arXiv.org/abs/0805.1573}{{\tt 0805.1573}}].

\bibitem{Grana:2003ek}
M.~Gra{\~{n}}a, T.~W. Grimm, H.~Jockers and J.~Louis,  {\em {Soft supersymmetry
  breaking in Calabi-Yau orientifolds with D-branes and fluxes}}, Nucl.Phys.
  {\bf B690} (2004) 21--61
[\href{http://www.arXiv.org/abs/hep-th/0312232}{{\tt hep-th/0312232}}].

\bibitem{Camara:2003ku}
P.~G. C{\'{a}}mara, L.~E. Ib{\'{a}}{\~{n}}ez and A.~M. Uranga,  {\em {Flux
  induced SUSY breaking soft terms}}, Nucl.Phys. {\bf B689} (2004) 195--242
[\href{http://www.arXiv.org/abs/hep-th/0311241}{{\tt hep-th/0311241}}].

\bibitem{Camara:2004jj}
P.~G. C{\'{a}}mara, L.~E. Ib{\'{a}}{\~{n}}ez and A.~M. Uranga,  {\em
  {Flux-induced SUSY-breaking soft terms on D7-D3 brane systems}}, Nucl.Phys.
  {\bf B708} (2005) 268--316
[\href{http://www.arXiv.org/abs/hep-th/0408036}{{\tt hep-th/0408036}}].

\bibitem{Lust:2004fi}
D.~L{\"{u}}st, S.~Reffert and S.~Stieberger,  {\em {Flux-induced soft
  supersymmetry breaking in chiral type IIB orientifolds with D3 / D7-branes}},
  Nucl.Phys. {\bf B706} (2005) 3--52
[\href{http://www.arXiv.org/abs/hep-th/0406092}{{\tt hep-th/0406092}}].

\bibitem{Jockers:2004yj}
H.~Jockers and J.~Louis,  {\em {The Effective action of D7-branes in N = 1
  Calabi-Yau orientifolds}}, Nucl.Phys. {\bf B705} (2005) 167--211
[\href{http://www.arXiv.org/abs/hep-th/0409098}{{\tt hep-th/0409098}}].

\bibitem{Jockers:2005zy}
H.~Jockers and J.~Louis,  {\em {D-terms and F-terms from D7-brane fluxes}},
  Nucl.Phys. {\bf B718} (2005) 203--246
[\href{http://www.arXiv.org/abs/hep-th/0502059}{{\tt hep-th/0502059}}].

\bibitem{Grimm:2004uq}
T.~W. Grimm and J.~Louis,  {\em {The Effective action of N = 1 Calabi-Yau
  orientifolds}}, Nucl.Phys. {\bf B699} (2004) 387--426
[\href{http://www.arXiv.org/abs/hep-th/0403067}{{\tt hep-th/0403067}}].

\bibitem{Grimm:2004ua}
T.~W. Grimm and J.~Louis,  {\em {The Effective action of type IIA Calabi-Yau
  orientifolds}}, Nucl.Phys. {\bf B718} (2005) 153--202
[\href{http://www.arXiv.org/abs/hep-th/0412277}{{\tt hep-th/0412277}}].

\bibitem{Grimm:2008dq}
T.~W. Grimm, T.-W. Ha, A.~Klemm and D.~Klevers,  {\em {The D5-brane effective
  action and superpotential in N=1 compactifications}}, Nucl.Phys. {\bf B816}
  (2009) 139--184
[\href{http://www.arXiv.org/abs/0811.2996}{{\tt 0811.2996}}].

\bibitem{Blumenhagen:2006ci}
R.~Blumenhagen, B.~K{\"{o}}rs, D.~L{\"{u}}st and S.~Stieberger,  {\em
  {Four-dimensional String Compactifications with D-Branes, Orientifolds and
  Fluxes}}, Phys.Rept. {\bf 445} (2007) 1--193
[\href{http://www.arXiv.org/abs/hep-th/0610327}{{\tt hep-th/0610327}}].

\bibitem{DeWolfe:2002nn}
O.~DeWolfe and S.~B. Giddings,  {\em {Scales and hierarchies in warped
  compactifications and brane worlds}}, Phys.Rev. {\bf D67} (2003) 066008
[\href{http://www.arXiv.org/abs/hep-th/0208123}{{\tt hep-th/0208123}}].

\bibitem{Giddings:2005ff}
S.~B. Giddings and A.~Maharana,  {\em {Dynamics of warped compactifications and
  the shape of the warped landscape}}, Phys. Rev. {\bf D73} (2006) 126003
[\href{http://www.arXiv.org/abs/hep-th/0507158}{{\tt hep-th/0507158}}].

\bibitem{Frey:2006wv}
A.~R. Frey and A.~Maharana,  {\em {Warped spectroscopy: Localization of frozen
  bulk modes}}, JHEP {\bf 08} (2006) 021
[\href{http://www.arXiv.org/abs/hep-th/0603233}{{\tt hep-th/0603233}}].

\bibitem{Burgess:2006mn}
C.~P. Burgess, P.~G. C{\'{a}}mara, S.~P. de~Alwis, S.~B. Giddings, A.~Maharana,
  F.~Quevedo and K.~Suruliz,  {\em {Warped Supersymmetry Breaking}}, JHEP {\bf
  0804} (2008) 053
[\href{http://www.arXiv.org/abs/hep-th/0610255}{{\tt hep-th/0610255}}].

\bibitem{Douglas:2007tu}
M.~R. Douglas, J.~Shelton and G.~Torroba,  {\em {Warping and supersymmetry
  breaking}},
\href{http://www.arXiv.org/abs/0704.4001}{{\tt 0704.4001}}.

\bibitem{Koerber:2007xk}
P.~Koerber and L.~Martucci,  {\em {From ten to four and back again: how to
  generalize the geometry}}, JHEP {\bf 08} (2007) 059
[\href{http://www.arXiv.org/abs/0707.1038}{{\tt 0707.1038}}].

\bibitem{Shiu:2008ry}
G.~Shiu, G.~Torroba, B.~Underwood and M.~R. Douglas,  {\em {Dynamics of Warped
  Flux Compactifications}}, JHEP {\bf 06} (2008) 024
[\href{http://www.arXiv.org/abs/0803.3068}{{\tt 0803.3068}}].

\bibitem{Douglas:2008jx}
M.~R. Douglas and G.~Torroba,  {\em {Kinetic terms in warped
  compactifications}}, JHEP {\bf 05} (2009) 013
[\href{http://www.arXiv.org/abs/0805.3700}{{\tt 0805.3700}}].

\bibitem{Frey:2008xw}
A.~R. Frey, G.~Torroba, B.~Underwood and M.~R. Douglas,  {\em {The Universal
  Kaehler Modulus in Warped Compactifications}}, JHEP {\bf 01} (2009) 036
[\href{http://www.arXiv.org/abs/0810.5768}{{\tt 0810.5768}}].

\bibitem{Marchesano:2008rg}
F.~Marchesano, P.~McGuirk and G.~Shiu,  {\em {Open String Wavefunctions in
  Warped Compactifications}}, JHEP {\bf 0904} (2009) 095
[\href{http://www.arXiv.org/abs/0812.2247}{{\tt 0812.2247}}].

\bibitem{Martucci:2009sf}
L.~Martucci,  {\em {On moduli and effective theory of N=1 warped flux
  compactifications}}, JHEP {\bf 05} (2009) 027
[\href{http://www.arXiv.org/abs/0902.4031}{{\tt 0902.4031}}].

\bibitem{Chen:2009zi}
H.-Y. Chen, Y.~Nakayama and G.~Shiu,  {\em {On D3-brane Dynamics at Strong
  Warping}}, Int.J.Mod.Phys. {\bf A25} (2010) 2493--2513
[\href{http://www.arXiv.org/abs/0905.4463}{{\tt 0905.4463}}].

\bibitem{Douglas:2009zn}
M.~R. Douglas,  {\em {Effective potential and warp factor dynamics}}, JHEP {\bf
  03} (2010) 071
[\href{http://www.arXiv.org/abs/0911.3378}{{\tt 0911.3378}}].

\bibitem{Douglas:2010rt}
M.~R. Douglas and R.~Kallosh,  {\em {Compactification on negatively curved
  manifolds}}, JHEP {\bf 1006} (2010) 004
[\href{http://www.arXiv.org/abs/1001.4008}{{\tt 1001.4008}}].

\bibitem{Blaback:2010sj}
J.~Bl{\r{a}}b{\"{a}}ck, U.~H. Danielsson, D.~Junghans, T.~Van~Riet, T.~Wrase
  and M.~Zagermann,  {\em {Smeared versus localised sources in flux
  compactifications}}, JHEP {\bf 1012} (2010) 043
[\href{http://www.arXiv.org/abs/1009.1877}{{\tt 1009.1877}}].

\bibitem{Underwood:2010pm}
B.~Underwood,  {\em {A Breathing Mode for Warped Compactifications}},
  Class.Quant.Grav. {\bf 28} (2011) 195013
[\href{http://www.arXiv.org/abs/1009.4200}{{\tt 1009.4200}}].

\bibitem{Marchesano:2010bs}
F.~Marchesano, P.~McGuirk and G.~Shiu,  {\em {Chiral matter wavefunctions in
  warped compactifications}}, JHEP {\bf 1105} (2011) 090
[\href{http://www.arXiv.org/abs/1012.2759}{{\tt 1012.2759}}].

\bibitem{Blaback:2012mu}
J.~Bl{\r{a}}b{\"{a}}ck, B.~Janssen, T.~Van~Riet and B.~Vercnocke,  {\em
  {Fractional branes, warped compactifications and backreacted orientifold
  planes}}, JHEP {\bf 1210} (2012) 139
[\href{http://www.arXiv.org/abs/1207.0814}{{\tt 1207.0814}}].

\bibitem{Frey:2013bha}
A.~R. Frey and J.~Roberts,  {\em {The Dimensional Reduction and Kähler Metric
  of Forms In Flux and Warping}}, JHEP {\bf 1310} (2013) 021
[\href{http://www.arXiv.org/abs/1308.0323}{{\tt 1308.0323}}].

\bibitem{Junghans:2014xfa}
D.~Junghans,  {\em {Dynamics of warped flux compactifications with backreacting
  anti-branes}},
\href{http://www.arXiv.org/abs/1402.4571}{{\tt 1402.4571}}.

\bibitem{Baumann:2006th}
D.~Baumann, A.~Dymarsky, I.~R. Klebanov, J.~M. Maldacena, L.~P. McAllister and
  A.~Murugan,  {\em {On D3-brane Potentials in Compactifications with Fluxes
  and Wrapped D-branes}}, JHEP {\bf 0611} (2006) 031
[\href{http://www.arXiv.org/abs/hep-th/0607050}{{\tt hep-th/0607050}}].

\bibitem{Kachru:2003sx}
S.~Kachru, R.~Kallosh, A.~D. Linde, J.~M. Maldacena, L.~P. McAllister and S.~P.
  Trivedi,  {\em {Towards inflation in string theory}}, JCAP {\bf 0310} (2003)
  013
[\href{http://www.arXiv.org/abs/hep-th/0308055}{{\tt hep-th/0308055}}].

\bibitem{Polchinski:1995mt}
J.~Polchinski,  {\em {Dirichlet Branes and Ramond-Ramond charges}},
  Phys.Rev.Lett. {\bf 75} (1995) 4724--4727
[\href{http://www.arXiv.org/abs/hep-th/9510017}{{\tt hep-th/9510017}}].

\bibitem{Witten:1995im}
E.~Witten,  {\em {Bound states of strings and p-branes}}, Nucl.Phys. {\bf B460}
  (1996) 335--350
[\href{http://www.arXiv.org/abs/hep-th/9510135}{{\tt hep-th/9510135}}].

\bibitem{Polchinski:1996na}
J.~Polchinski,  {\em {Tasi lectures on D-branes}},
\href{http://www.arXiv.org/abs/hep-th/9611050}{{\tt hep-th/9611050}}.

\bibitem{Myers:1999ps}
R.~C. Myers,  {\em {Dielectric branes}}, JHEP {\bf 9912} (1999) 022
[\href{http://www.arXiv.org/abs/hep-th/9910053}{{\tt hep-th/9910053}}].

\bibitem{Bershadsky:1995qy}
M.~Bershadsky, C.~Vafa and V.~Sadov,  {\em {D-branes and topological field
  theories}}, Nucl.Phys. {\bf B463} (1996) 420--434
[\href{http://www.arXiv.org/abs/hep-th/9511222}{{\tt hep-th/9511222}}].

\bibitem{Green:1996dd}
M.~B. Green, J.~A. Harvey and G.~W. Moore,  {\em {I-brane inflow and anomalous
  couplings on d-branes}}, Class.Quant.Grav. {\bf 14} (1997) 47--52
[\href{http://www.arXiv.org/abs/hep-th/9605033}{{\tt hep-th/9605033}}].

\bibitem{Cheung:1997az}
Y.-K.~E. Cheung and Z.~Yin,  {\em {Anomalies, branes, and currents}},
  Nucl.Phys. {\bf B517} (1998) 69--91
[\href{http://www.arXiv.org/abs/hep-th/9710206}{{\tt hep-th/9710206}}].

\bibitem{Minasian:1997mm}
R.~Minasian and G.~W. Moore,  {\em {K theory and Ramond-Ramond charge}}, JHEP
  {\bf 9711} (1997) 002
[\href{http://www.arXiv.org/abs/hep-th/9710230}{{\tt hep-th/9710230}}].

\bibitem{Fotopoulos:2001pt}
A.~Fotopoulos,  {\em {On (alpha-prime)**2 corrections to the D-brane action for
  nongeodesic world volume embeddings}}, JHEP {\bf 0109} (2001) 005
[\href{http://www.arXiv.org/abs/hep-th/0104146}{{\tt hep-th/0104146}}].

\bibitem{Wyllard:2001ye}
N.~Wyllard,  {\em {Derivative corrections to the D-brane Born-Infeld action:
  Nongeodesic embeddings and the Seiberg-Witten map}}, JHEP {\bf 0108} (2001)
  027
[\href{http://www.arXiv.org/abs/hep-th/0107185}{{\tt hep-th/0107185}}].

\bibitem{Fotopoulos:2002wy}
A.~Fotopoulos and A.~A. Tseytlin,  {\em {On gravitational couplings in D-brane
  action}}, JHEP {\bf 0212} (2002) 001
[\href{http://www.arXiv.org/abs/hep-th/0211101}{{\tt hep-th/0211101}}].

\bibitem{Morales:1998ux}
J.~F. Morales, C.~A. Scrucca and M.~Serone,  {\em {Anomalous couplings for
  D-branes and O-planes}}, Nucl.Phys. {\bf B552} (1999) 291--315
[\href{http://www.arXiv.org/abs/hep-th/9812071}{{\tt hep-th/9812071}}].

\bibitem{Stefanski:1998he}
B.~Stefanski,  {\em {Gravitational couplings of D-branes and O-planes}},
  Nucl.Phys. {\bf B548} (1999) 275--290
[\href{http://www.arXiv.org/abs/hep-th/9812088}{{\tt hep-th/9812088}}].

\bibitem{Craps:1998fn}
B.~Craps and F.~Roose,  {\em {Anomalous D-brane and orientifold couplings from
  the boundary state}}, Phys.Lett. {\bf B445} (1998) 150--159
[\href{http://www.arXiv.org/abs/hep-th/9808074}{{\tt hep-th/9808074}}].

\bibitem{Craps:1998tw}
B.~Craps and F.~Roose,  {\em {(Non)anomalous D-brane and O-plane couplings: The
  Normal bundle}}, Phys.Lett. {\bf B450} (1999) 358
[\href{http://www.arXiv.org/abs/hep-th/9812149}{{\tt hep-th/9812149}}].

\bibitem{Scrucca:1999uz}
C.~A. Scrucca and M.~Serone,  {\em {Anomalies and inflow on D-branes and O -
  planes}}, Nucl.Phys. {\bf B556} (1999) 197--221
[\href{http://www.arXiv.org/abs/hep-th/9903145}{{\tt hep-th/9903145}}].

\bibitem{Dasgupta:1997cd}
K.~Dasgupta, D.~P. Jatkar and S.~Mukhi,  {\em {Gravitational couplings and Z(2)
  orientifolds}}, Nucl.Phys. {\bf B523} (1998) 465--484
[\href{http://www.arXiv.org/abs/hep-th/9707224}{{\tt hep-th/9707224}}].

\bibitem{Dasgupta:1997wd}
K.~Dasgupta and S.~Mukhi,  {\em {Anomaly inflow on orientifold planes}}, JHEP
  {\bf 9803} (1998) 004
[\href{http://www.arXiv.org/abs/hep-th/9709219}{{\tt hep-th/9709219}}].

\bibitem{Schnitzer:2002rt}
H.~J. Schnitzer and N.~Wyllard,  {\em {An Orientifold of AdS(5) x T**11 with
  D7-branes, the associated alpha-prime**2 corrections and their role in the
  dual N=1 Sp(2N + 2M) x Sp(2N) gauge theory}}, JHEP {\bf 0208} (2002) 012
[\href{http://www.arXiv.org/abs/hep-th/0206071}{{\tt hep-th/0206071}}].

\bibitem{Giddings:2001yu}
S.~B. Giddings, S.~Kachru and J.~Polchinski,  {\em {Hierarchies from fluxes in
  string compactifications}}, Phys.Rev. {\bf D66} (2002) 106006
[\href{http://www.arXiv.org/abs/hep-th/0105097}{{\tt hep-th/0105097}}].

\bibitem{Garousi:2006zh}
M.~R. Garousi,  {\em {Superstring scattering from O-planes}}, Nucl.Phys. {\bf
  B765} (2007) 166--184
[\href{http://www.arXiv.org/abs/hep-th/0611173}{{\tt hep-th/0611173}}].

\bibitem{Robbins:2014ara}
D.~Robbins and Z.~Wang,  {\em {Higher Derivative Corrections to O-plane
  Actions: NS-NS Sector}},
\href{http://www.arXiv.org/abs/1401.4180}{{\tt 1401.4180}}.

\bibitem{Nakahara:2003nw}
M.~Nakahara, {\em {Geometry, topology and physics}}.
\newblock Taylor \& Francis,
2003.
\newblock

\bibitem{Buscher:1987sk}
T.~H. Buscher,  {\em {A Symmetry of the String Background Field Equations}},
  Phys.Lett. {\bf B194} (1987)
59.

\bibitem{Wyllard:2000qe}
N.~Wyllard,  {\em {Derivative corrections to D-brane actions with constant
  background fields}}, Nucl.Phys. {\bf B598} (2001) 247--275
[\href{http://www.arXiv.org/abs/hep-th/0008125}{{\tt hep-th/0008125}}].

\bibitem{Wijnholt:2003pw}
M.~Wijnholt,  {\em {On curvature squared corrections for D-brane actions}},
\href{http://www.arXiv.org/abs/hep-th/0301029}{{\tt hep-th/0301029}}.

\bibitem{Becker:2010ij}
K.~Becker, G.~Guo and D.~Robbins,  {\em {Higher Derivative Brane Couplings from
  T-Duality}}, JHEP {\bf 1009} (2010) 029
[\href{http://www.arXiv.org/abs/1007.0441}{{\tt 1007.0441}}].

\bibitem{Becker:2011ar}
K.~Becker, G.~Guo and D.~Robbins,  {\em {Four-Derivative Brane Couplings from
  String Amplitudes}}, JHEP {\bf 1112} (2011) 050
[\href{http://www.arXiv.org/abs/1110.3831}{{\tt 1110.3831}}].

\bibitem{Garousi:2009dj}
M.~R. Garousi,  {\em {T-duality of Curvature terms in D-brane actions}}, JHEP
  {\bf 1002} (2010) 002
[\href{http://www.arXiv.org/abs/0911.0255}{{\tt 0911.0255}}].

\bibitem{Garousi:2010ki}
M.~R. Garousi,  {\em {Ramond-Ramond field strength couplings on D-branes}},
  JHEP {\bf 1003} (2010) 126
[\href{http://www.arXiv.org/abs/1002.0903}{{\tt 1002.0903}}].

\bibitem{Garousi:2010rn}
M.~R. Garousi,  {\em {T-duality of anomalous Chern-Simons couplings}},
  Nucl.Phys. {\bf B852} (2011) 320--335
[\href{http://www.arXiv.org/abs/1007.2118}{{\tt 1007.2118}}].

\bibitem{Garousi:2010bm}
M.~R. Garousi and M.~Mir,  {\em {On RR couplings on D-branes at order
  $O(\alpha'^2)$}}, JHEP {\bf 1102} (2011) 008
[\href{http://www.arXiv.org/abs/1012.2747}{{\tt 1012.2747}}].

\bibitem{Garousi:2011ut}
M.~R. Garousi and M.~Mir,  {\em {Towards extending the Chern-Simons couplings
  at order $O(\alpha'^2)$}}, JHEP {\bf 1105} (2011) 066
[\href{http://www.arXiv.org/abs/1102.5510}{{\tt 1102.5510}}].

\bibitem{Garousi:2011fc}
M.~R. Garousi,  {\em {S-duality of D-brane action at order $O(\alpha'^2)$}},
  Phys.Lett. {\bf B701} (2011) 465--470
[\href{http://www.arXiv.org/abs/1103.3121}{{\tt 1103.3121}}].

\bibitem{Velni:2012sv}
K.~B. Velni and M.~R. Garousi,  {\em {S-matrix elements from T-duality}},
  Nucl.Phys. {\bf B869} (2013) 216--241
[\href{http://www.arXiv.org/abs/1204.4978}{{\tt 1204.4978}}].

\bibitem{Velni:2013jha}
K.~B. Velni and M.~R. Garousi,  {\em {Ramond-Ramond S-matrix elements from
  T-dual Ward identity}},
\href{http://www.arXiv.org/abs/1312.0213}{{\tt 1312.0213}}.

\bibitem{Hatefi:2010ik}
E.~Hatefi,  {\em {On effective actions of BPS branes and their higher
  derivative corrections}}, JHEP {\bf 1005} (2010) 080
[\href{http://www.arXiv.org/abs/1003.0314}{{\tt 1003.0314}}].

\bibitem{Hatefi:2012ve}
E.~Hatefi and I.~Park,  {\em {More on closed string induced higher derivative
  interactions on D-branes}}, Phys.Rev. {\bf D85} (2012) 125039
[\href{http://www.arXiv.org/abs/1203.5553}{{\tt 1203.5553}}].

\bibitem{Hatefi:2012zh}
E.~Hatefi,  {\em {Shedding light on new Wess-Zumino couplings with their
  corrections to all orders in alpha-prime}}, JHEP {\bf 1304} (2013) 070
[\href{http://www.arXiv.org/abs/1211.2413}{{\tt 1211.2413}}].

\bibitem{Gross:1986iv}
D.~J. Gross and E.~Witten,  {\em {Superstring Modifications of Einstein's
  Equations}}, Nucl.Phys. {\bf B277} (1986)
1.

\bibitem{Tseytlin:1986zz}
A.~A. Tseytlin,  {\em {Ambiguity in the Effective Action in String Theories}},
  Phys.Lett. {\bf B176} (1986)
92.

\bibitem{Tseytlin:1993df}
A.~A. Tseytlin,  {\em {On field redefinitions and exact solutions in string
  theory}}, Phys.Lett. {\bf B317} (1993) 559--564
[\href{http://www.arXiv.org/abs/hep-th/9308042}{{\tt hep-th/9308042}}].

\bibitem{Forger:1996vj}
K.~F{\"{o}}rger, B.~A. Ovrut, S.~J. Theisen and D.~Waldram,  {\em {Higher
  derivative gravity in string theory}}, Phys.Lett. {\bf B388} (1996) 512--520
[\href{http://www.arXiv.org/abs/hep-th/9605145}{{\tt hep-th/9605145}}].

\bibitem{Peeters:2007wn}
K.~Peeters,  {\em {Introducing Cadabra: A Symbolic computer algebra system for
  field theory problems}},
\href{http://www.arXiv.org/abs/hep-th/0701238}{{\tt hep-th/0701238}}.

\bibitem{Peeters2}
K.~Peeters,  {\em {A field-theory motivated approach to symbolic computer
  algebra}}, Comp. Phys. Commun. {\bf 176} (2006) 550--558
  [\href{http://www.arXiv.org/abs/cs.sc/0608005}{{\tt cs.sc/0608005}}].

\bibitem{Becker:1996gj}
K.~Becker and M.~Becker,  {\em {M theory on eight manifolds}}, Nucl.Phys. {\bf
  B477} (1996) 155--167
[\href{http://www.arXiv.org/abs/hep-th/9605053}{{\tt hep-th/9605053}}].

\bibitem{Dasgupta:1999ss}
K.~Dasgupta, G.~Rajesh and S.~Sethi,  {\em {M theory, orientifolds and G -
  flux}}, JHEP {\bf 9908} (1999) 023
[\href{http://www.arXiv.org/abs/hep-th/9908088}{{\tt hep-th/9908088}}].

\bibitem{Gukov:1999ya}
S.~Gukov, C.~Vafa and E.~Witten,  {\em {CFT's from Calabi-Yau four folds}},
  Nucl.Phys. {\bf B584} (2000) 69--108
[\href{http://www.arXiv.org/abs/hep-th/9906070}{{\tt hep-th/9906070}}].

\bibitem{Greene:2000gh}
B.~R. Greene, K.~Schalm and G.~Shiu,  {\em {Warped compactifications in M and F
  theory}}, Nucl.Phys. {\bf B584} (2000) 480--508
[\href{http://www.arXiv.org/abs/hep-th/0004103}{{\tt hep-th/0004103}}].

\bibitem{Wald:1984rg}
R.~M. Wald, {\em {General Relativity}}.
\newblock The University of Chicago Press,
1984.
\newblock

\bibitem{Kakushadze:2001bd}
Z.~Kakushadze,  {\em {Orientiworld}}, JHEP {\bf 0110} (2001) 031
[\href{http://www.arXiv.org/abs/hep-th/0109054}{{\tt hep-th/0109054}}].

\bibitem{Jockers:2005pn}
H.~Jockers,  {\em {The Effective action of D-branes in Calabi-Yau orientifold
  compactifications}}, Fortsch.Phys. {\bf 53} (2005) 1087--1175
[\href{http://www.arXiv.org/abs/hep-th/0507042}{{\tt hep-th/0507042}}].

\bibitem{Frey:2002hf}
A.~R. Frey and J.~Polchinski,  {\em {N=3 warped compactifications}}, Phys.Rev.
  {\bf D65} (2002) 126009
[\href{http://www.arXiv.org/abs/hep-th/0201029}{{\tt hep-th/0201029}}].

\bibitem{Braun:2008ua}
A.~P. Braun, A.~Hebecker and H.~Triendl,  {\em {D7-Brane Motion from M-Theory
  Cycles and Obstructions in the Weak Coupling Limit}}, Nucl.Phys. {\bf B800}
  (2008) 298--329
[\href{http://www.arXiv.org/abs/0801.2163}{{\tt 0801.2163}}].

\bibitem{Sethi:1996es}
S.~Sethi, C.~Vafa and E.~Witten,  {\em {Constraints on low dimensional string
  compactifications}}, Nucl.Phys. {\bf B480} (1996) 213--224
[\href{http://www.arXiv.org/abs/hep-th/9606122}{{\tt hep-th/9606122}}].

\bibitem{Klemm:1996ts}
A.~Klemm, B.~Lian, S.~S. Roan and S.-T. Yau,  {\em {Calabi-Yau fourfolds for M
  theory and F theory compactifications}}, Nucl.Phys. {\bf B518} (1998)
  515--574
[\href{http://www.arXiv.org/abs/hep-th/9701023}{{\tt hep-th/9701023}}].

\bibitem{Beasley:2008dc}
C.~Beasley, J.~J. Heckman and C.~Vafa,  {\em {GUTs and Exceptional Branes in
  F-theory - I}}, JHEP {\bf 0901} (2009) 058
[\href{http://www.arXiv.org/abs/0802.3391}{{\tt 0802.3391}}].

\bibitem{Donagi:2008ca}
R.~Donagi and M.~Wijnholt,  {\em {Model Building with F-Theory}},
  Adv.Theor.Math.Phys. {\bf 15} (2011) 1237--1318
[\href{http://www.arXiv.org/abs/0802.2969}{{\tt 0802.2969}}].

\bibitem{future}
W.~Cottrell, D.~Junghans and G.~Shiu,  to appear.

\bibitem{Sen:2013oza}
A.~Sen,  {\em {S-duality Improved Superstring Perturbation Theory}},
\href{http://www.arXiv.org/abs/1304.0458}{{\tt 1304.0458}}.

\end{thebibliography}\endgroup

\end{document}